\documentclass[12pt]{article} 
\usepackage{graphicx,subfigure,latexsym}
\usepackage{amsmath,amssymb,amsfonts}
\usepackage{bm}

\newcommand{\be}{\begin{equation}}
\newcommand{\ee}{\end{equation}}
\newcommand{\bear}{\begin{eqnarray}}
\newcommand{\eear}{\end{eqnarray}}
\newcommand{\ba}{\begin{array}}
\newcommand{\ea}{\end{array}}

\def \be {\begin{equation}}
\def \ee {\end{equation}}

\def \bes {\begin{subequations}}
\def \ees {\end{subequations}}

\def \<{\langle}
\def \>{\rangle}
\def \+{\dagger}
\def \({\left(}
\def \){\right)}
\def \[{\left[}
\def \]{\right]}

\begin{document}

\begin{titlepage}
\vfill
\begin{flushright}
{\normalsize RBRC-1184}\\
\end{flushright}

\vfill
\begin{center}
{\Large\bf  Jet quenching parameter of quark-gluon plasma in strong magnetic field:  Perturbative QCD and AdS/CFT correspondence}

\vskip 0.3in

\vskip 0.3in
Shiyong Li$^{1}$\footnote{e-mail: {\tt sli72@uic.edu}}, Kiminad A. Mamo$^{1}$\footnote{e-mail: {\tt kabebe2@uic.edu}} and
Ho-Ung Yee$^{1,2}$\footnote{e-mail: {\tt hyee@uic.edu}}
\vskip 0.15in

{\it $^{1}$ Department of Physics, University of Illinois, Chicago, Illinois 60607}\\[0.15in]
{\it $^{2}$ RIKEN-BNL Research Center, Brookhaven National Laboratory,}\\
{\it Upton, New York
11973-5000}\\[0.15in]
{\normalsize  2016}

\end{center}

\vfill

\begin{abstract}

We compute the jet quenching parameter $\hat q$ of QCD plasma in the presence of strong magnetic field in both weakly and strongly
coupled regimes. In weakly coupled regime, we compute $\hat q$ in perturbative QCD at complete leading order (that is, leading log as well as the constant under the log) in QCD coupling constant $\alpha_s$, assuming the hierarchy of scales $\alpha_s eB\ll T^2\ll eB$.
We consider two cases of jet orientations with respect to the magnetic field: 1) the case of jet moving parallel to the magnetic field, 2) the case jet moving perpendicular to the magnetic field. In the former case, we find $\hat q\sim \alpha_s^2 (eB)T\log(1/\alpha_s)$, while in the latter we have $\hat q\sim \alpha_s^2 (eB)T\log(T^2/\alpha_seB)$. In both cases, this leading order result arises from the scatterings with thermally populated lowest Landau level quarks. In strongly coupled regime described by AdS/CFT correspondence,
we find $\hat q\sim \sqrt{\lambda}(eB)T$ or $\hat q\sim \sqrt{\lambda}\sqrt{eB}T^2$ in the same hierarchy of $T^2\ll eB$ depending on whether the jet is moving parallel or perpendicular to the magnetic field, respectively, which indicates a universal dependence of $\hat q$ on $(eB)T$ in both regimes for the parallel case, the origin of which should be the transverse density of lowest Landau level states proportional to $eB$.
Finally, the asymmetric transverse momentum diffusion in the case of jet moving perpendicular to the magnetic field may give an interesting azimuthal asymmetry of the gluon Bremsstrahlung spectrum in the BDMPS-Z formalism.

\end{abstract}

\vfill

\end{titlepage}
\setcounter{footnote}{0}

\baselineskip 18pt \pagebreak
\renewcommand{\thepage}{\arabic{page}}
\pagebreak

\section{Introduction}

The energy loss of a high energy jet in the QCD plasma via gluon Bremstrahlung, described by BDMPS-Z formalism in large scattering number limit \cite{Baier:1996kr,Baier:1996sk,Zakharov:1997uu,Gyulassy:2000er,Salgado:2003gb}, rests on a single parameter $\hat q$, the jet quenching parameter. It is defined as the transverse momentum diffusion constant of the (emitted) gluon per unit length of the jet trajectory: $\hat q=\langle \bm p_\perp^2\rangle/dz$ \cite{Baier:1996sk}. In our computation, we will call any fast moving color charged object with some representation $R$ a jet, since in the eikonal limit the identity of the object should not matter except its color charge (this includes the emitted gluon as well).
The same parameter also gives the damping rate of an energetic small dipole of size $b$ by $\Gamma^{\rm dipole}={1\over 2}\hat q b^2$ in small $b$ limit.
This connection between the two can be understood as follows.
The amplitude square of the gluon Bremstrahlung is a product of transition amplitude forward in time and its complex conjugate.
The conjugate amplitude can be put as
\be
\left(\langle f|U(t)|i\rangle\right)^*=\langle \bar f|(U(t))^*|\bar i\rangle\,,
\ee where $U(t)$ is the time-evolution operator and $|\bar i\rangle$ is a time-inversion state of the initial state $|i\rangle$, which in Schrodinger picture is just the complex conjugate wave function of the original wave-function. Since the time-inverse operator $U(t)^*$ describes a negative energy state with opposite color charge,
the complex conjugate of transition amplitude can be put as an ordinary transition amplitude
of a jet, but with a negative energy and opposite color charge, which evolves with time-reversed propagator $U(t)^*$. Let's call this ''anti-jet''. This is nothing but the evolution on the second contour in Schwinger-Keldysh formalism for complex conjugate amplitudes. The key element is that the thermally fluctuating soft gauge fields that are the main source of scatterings with the jet are classical fields in nature, which are ''r''-type fields in the language of Schwinger-Keldysh formalism: these classical soft r-type fluctuations give leading order contributions to the total scattering rate to the jet, due to Bose-Einstein enhancement in the soft region, $n_B(\omega)\sim T/\omega$ for $\omega\ll T$.
As these r-type fields have the same values on both contours in the Schwinger-Keldysh formalism, it doesn't matter on which contour we put the anti-jet for the computation of soft scatterings with them.
If we choose to put the jet and anti-jet together, they look just like a color dipole.
In BDMPS-Z formalism, we have jet-antijet-gluon three body system during the virtual process, which can be thought of as a collection of three color dipoles. The only difference between this jet-antijet pair and a real color dipole
is that the anti-jet has a negative kinetic energy: the damping rate part of the hamiltonian (i.e. the imaginary part) coming from soft scatterings with thermal fluctuations is the same between the two, since these scatterings care only the color charges of the pair. In large scattering number limit, the small size regime dominates, and the scattering amplitude becomes
\be
{\cal A}^{\rm pair}=(1-e^{i\bm b\cdot\bm q_\perp}){\cal A}^{\rm single}\approx -i(\bm b\cdot\bm q_\perp) {\cal A}^{\rm single},
\ee
where ${\cal A}_{\rm single}$ is the scattering amplitude with a single jet, $\bm q$ is the spatial part of the exchanged momentum, and $\bm b$ is the transverse size of the color dipole. This gives the damping rate part being
\be
\Gamma^{\rm pair}=\Gamma^{\rm dipole}\approx \int d^3\bm q {d\Gamma^{\rm single}\over d^3\bm q}(\bm b\cdot\bm q_\perp)^2={1\over 2}b^2\int
d^3\bm q {d\Gamma^{\rm single}\over d^3\bm q} \bm q_\perp^2={1\over 2}b^2 \hat q\,,
\ee
with the conventional definition of $\hat q$ being the transverse momentum diffusion rate of a single jet.

We compute $\hat q$ in the presence of strong magnetic field limit $eB\gg T^2$, in both weakly coupled regime at leading order in $\alpha_s$ as well as in strongly coupled regime described by AdS/CFT correspondence. In the former case, we additionally assume $\alpha_s eB\ll T^2$, so that self-energy corrections from lowest Landau level states (LLL) of quarks to the ``hard'' particles of typical momenta $T$ can be neglected (see later sections for more details). Only with this additional assumption of small enough coupling $\alpha_s$, a systematic power counting scheme at weak coupling we employ can apply: this scheme was recently introduced in Ref.\cite{Fukushima:2015wck} to compute heavy-quark diffusion constant in strong magnetic field in perturbative QCD (pQCD). We follow the same scheme in this work.
We further neglect small quark mass corrections treating them massless: this is well-justified practically, $m_q^2/eB$ or $m_q^2/T^2$ is about  $10^{-4}$ for $T\sim 300$ MeV. In both weakly and strongly coupled regimes, we consider the two cases of jet motions: the jet moving parallel to the magnetic field and the one moving perpendicular to the magnetic field.

\section{Jet quenching in weakly coupled regime}

The leading order computation of $\hat q$ in small $\alpha_s$ can be done by first computing the scattering rate per unit momentum transfer,
$
{d\Gamma^{\rm single}/ d^3\bm q}$, from leading t-channel gluon exchange between hard thermal quarks or gluons and the jet.
Then the jet quenching parameter is computed as
\be
\hat q={1\over v}\int d\bm q^3 {d\Gamma^{\rm single}\over d^3\bm q} \bm q_\perp^2\,,
\ee
where $\bm q_\perp$ is the transverse component of the momentum transfer, and $1/v$ factor is from the translation between the diffusion constants ``per unit length'' and ``per unit time"~:~$d/dz=(1/v)d/dt$.
In the large jet momentum limit $P\gg T$, which is the case for either heavy-quarks ($P^0=M_Q\gg T$) or for a ultra-relativistic jet ($P\approx E(1,\bm v)$ with $E\gg T$ and $v\approx 1$), the leading power of $P$ in the Feynman diagrams arises only in the t-channel exchange diagrams. For the case of scatterings with thermal gluons, this statement is not gauge-invariant, but is true in the gauge $\epsilon\cdot P=\tilde\epsilon\cdot P=0$ where $\epsilon,\tilde\epsilon$ are polarizations of incoming and out-going gluons \cite{Moore:2004tg}. For a ultra-relativistic jet where $P$ is nearly light-like, this gauge is essentially the light-cone gauge.

The t-channel momentum exchange $\bm q$ involves a soft scale ($Q\ll T$) for leading log contributions (as we will see), which features logarithmic IR singularity for $\hat q$. This is cured by gluon self-energy corrections either from thermally excited LLL quarks or from thermally excited hard gluons.
Both give the screening masses for t-channel gluon exchange, the former being $m_{D,B}^2\sim \alpha_s eB$ and the latter $m_D^2\sim \alpha_s T^2$.
Under our assumption of $eB\gg T^2$, we can keep only the former Debye screening from the LLL states. We emphasize that the t-channel exchanged gluons for which we include the self-energy are space-like and soft.

On the other hand, the dispersion relations of scattering hard quarks and hard gluons generally get thermal mass corrections from the same self-energy but evaluated in nearly on-shell kinematic regions. They are of the same order, $\alpha_s eB$ or $\alpha_s T^2$. As our further assumption of $\alpha_s eB\ll T^2$, and hard quarks and gluons have typical momenta $T$, we can neglect the self-energy (i.e. thermal mass) for these scattering hard thermal particles in leading order computation: the leading order $\hat q$ comes from the hard momentum ($\sim T$) region of scattering particles.
These hard particles are then free particles in leading order treatment. In turn, this also justifies the computation of self-energy itself from 1-loop of hard particles in the loop: these hard particles in the loop are free particles, their thermal mass corrections give only higher order corrections to the self-energy. This leading order treatment is then self-consistent \cite{Fukushima:2015wck}.

We give a brief summary of results we will obtain in the next subsections of detailed computation of $\hat q$.
For the case of scattering with thermal gluons, due to an issue of gauge invariance that we mentioned above, one needs to work directly with this
formula computing somewhat challenging phase space integrals as done originally in Ref.\cite{Moore:2004tg}. The leading log contribution is however manageable,
which will be shown in the Appendix to be $\hat q_{\rm gluon}\sim \alpha_s^2 T^3\log\left(T^2\over\alpha_s eB\right)$.
On the other hand, the contribution coming from scatterings with LLL quarks will be shown to be
$\hat q_{\rm quarks}\sim \alpha_s^2 eB T \log\left(1\over\alpha_s\right)$, which is larger than $\hat q_{\rm gluon}$ by a factor of $eB/T^2\gg 1$.
The origin of this enhancement is basically the large density of states of LLL quarks which scales linearly with $eB T$ ($eB$ from the density of states of LLL in two transverse dimensions and $T$ from the longitudinal thermal distribution), while the density of states of gluons with thermal distribution scales only with $T^3$.
Therefore, the leading order $\hat q$ is provided by the scatterings with the thermally excited LLL quarks.

\begin{figure}[t]
 \centering
 \includegraphics[height=5.5cm]{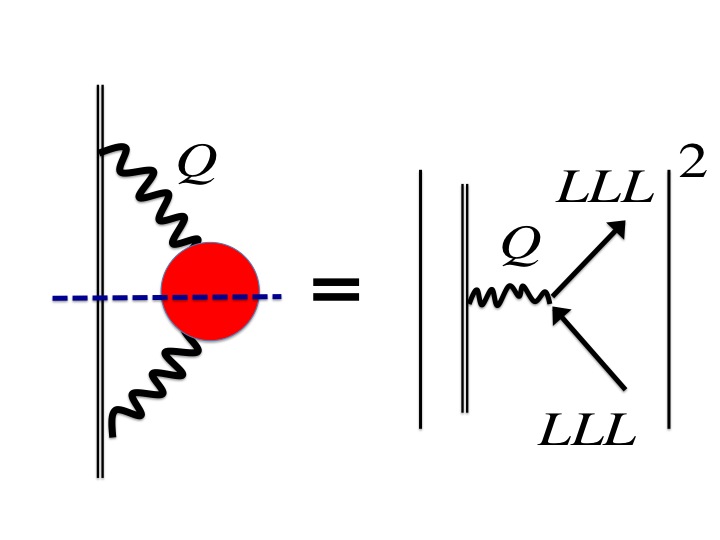}\caption{The imaginary cut of the jet self-energy is equal to the damping rate, that is, the total scattering rate with thermal (hard) particles, especially the lowest Landau level quarks. The exchanged gluon line is Debye screened by the same hard LLL states. \label{fig1}}
 \end{figure}

The t-channel process with LLL quarks is free of gauge-invariance issue, and in this case one can explore an alternative way of computing the t-channel scattering rate $d\Gamma/d^3\bm q$ from cutting the 1-loop retarded jet self-energy diagram, which gives the imaginary part of retarded jet self-energy or the damping rate of the jet,
\be
-{\rm Im}[\Sigma^R(P)]\sim \Gamma^{\rm single}=\int d^3\bm q {d\Gamma^{\rm single}\over d^3\bm q}\,,
\ee
where $\bm q$ is nothing but the loop momentum of the gluon line in the jet self-energy computation, and $\Sigma^R(P)$ is the retarded jet self-energy: see Figure \ref{fig1}.
 The internal gluon line should include its own self-energy coming from 1-loop hard thermal LLL states: that would be the Hard Thermal Loop propagator in the soft t-channel momentum region of $\bm q$, but now from the LLL states instead of more conventional free hard fermions/gluons. As argued in the above and in the Appendix, the contributions from hard gluons to this t-channel gluon self-energy is subdominant and neglected. Once we compute ${d\Gamma^{\rm single}/ d^3\bm q}$ in this method, we can compute $\hat q$ by weighting the integral by an additional factor of $\bm q_\perp^2$. This method seems much simpler, so we will adopt it in the next subsections.

\subsection{Scattering rate of the jet from its 1-loop self-energy}

For definiteness we assume that the jet is a fast moving fermion with momentum $P$, but the result in high $P$ limit is independent of this detail, due to eikonal reduction of jet propagation when $P\gg Q$: the only important fact is that the current of the jet in relativistic normalization is
\be
\bar U(P+Q)\gamma^\mu t^a U(P)\approx 2 P^\mu t^a\,,
\ee
where $t^a$ is the color charge of the jet.

The 1-loop retarded jet self-energy is given by $\Sigma^R(P)=(-i)\Sigma^{ra}(P)$ with ''ra''-self-energy in real-time formalism is
\be
\Sigma^{ra}(P)=(ig)^2 C_2^J \gamma^\beta\int {d^4 Q\over (2\pi)^4}\left[G^{rr}_{\alpha\beta}(Q) S^{ra}_{(0)}(P+Q)+G^{ar}_{\alpha\beta}(Q)S^{rr}_{(0)}(P+Q)\right] \gamma^\alpha\,,\label{selfE}
\ee
where $C_2^J$ is the color Casimir of the jet, and $G_{\alpha\beta}(Q)=\langle A_\alpha(Q) A_\beta(-Q)\rangle$ are the real-time gluon propagators without colors (or the color diagonal part defined by $\langle A^a_\alpha(Q) A^b_\beta(-Q)\rangle\equiv G_{\alpha\beta}(Q)\delta^{ab}$), and
$S_{(0)}(Q)$ are the bare jet propagator given by\footnote{In general, $S_{(0)}(Q)$ can be written as a sum of particle branch with a positive energy pole and anti-particle branch with a negative energy pole. We choose only the particle branch part, since the anti-particle branch decouples in high energy limit.}
\bear
S^{ra}_{(0)}(Q)&=&(-i){\gamma^0 {\cal P}(\bm q)\over q^0-\sqrt{\bm q^2+M^2}+i\epsilon}\,,\quad
S^{ar}_{(0)}(Q)=(-i){\gamma^0 {\cal P}(\bm q)\over q^0-\sqrt{\bm q^2+M^2}-i\epsilon}\,,\nonumber\\
S^{rr}_{(0)}(Q)&=&-\left({1\over 2}-n_F(q^0)\right)(2\pi)\gamma^0 {\cal P}(\bm q)\delta\left(q^0-\sqrt{\bm q^2+M^2}\right)\,,\label{fKMS}
\eear
with the spinor projection operator
\be
{\cal P}(\bm q)={1\over 2}\left({\bf 1}+{\gamma^0(\bm\gamma\cdot\bm q-iM)\over\sqrt{\bm q^2+M^2}}\right)\,,
\ee
and $M$ is the rest mass of the jet. We will consider relativistic cases where the jet momentum $\bm p\gg M$. Our metric convention in this work is $\eta=(-,+,+,+)$.
The self-energy re-summed jet propagator $S(P)$ is given by
\be
(S^{ra}(P))^{-1}=(S^{ra}_{(0)}(P))^{-1}-\Sigma^{ra}(P)\,,\label{resumpr}
\ee
and the damping rate of the jet $\Gamma^{\rm single}$ is identified by the ansatz
\be
S^{ra}(P)\approx (-i){\gamma^0 {\cal P}(\bm p)\over p^0-\sqrt{\bm p^2+M^2}+i\Gamma^{\rm single}/2}
\ee
neglecting a mass shift and wave function renormalization which are from the real part of $\Sigma^R(P)$ instead of the imaginary part.
This ansatz is equivalent to
\be
(S^{ra}(P))^{-1}\approx (-i) {\cal P}(\bm p)\gamma^0\left(p^0-\sqrt{\bm p^2+M^2}+i\Gamma^{\rm single}/2\right)= (S^{ra}_{(0)}(P))^{-1}
+ {\cal P}(\bm p)\gamma^0\Gamma^{\rm single}/2\,,
\ee
and comparing with (\ref{resumpr}) and using ${\rm Tr}({\cal P}(\bm p))=2$, we have
\be
\Gamma^{\rm single}={\rm Re}\left[{\rm Tr}\left(\Sigma^{ra}(P)\gamma^0 {\cal P}(\bm p)\right)\right]\bigg|_{p^0=\sqrt{{\bm p}^2+M^2}}=-{\rm Im}\left[{\rm Tr}\left(\Sigma^R(P)\gamma^0 {\cal P}(\bm p)\right)\right]\bigg|_{p^0=\sqrt{{\bm p}^2+M^2}}\,,
\ee
which is the desired formula relating the damping rate of the jet with the imaginary part of its retarded self-energy.

Using the explicit expression (\ref{selfE}) for $\Sigma^{ra}(P)$, and (\ref{fKMS}), and the similar thermal relations for gluon propagators
\bear
G^{ar}_{\alpha\beta}(Q)&=&\left(G^{ra}_{\beta\alpha}(Q)\right)^*\,,\nonumber\\
G^{rr}_{\alpha\beta}(Q)&=&\left({1\over 2}+n_B(q^0)\right)\left(G^{ra}_{\alpha\beta}(Q)-G^{ar}_{\alpha\beta}(Q)\right)\equiv
\left({1\over 2}+n_B(q^0)\right)\rho_{\alpha\beta}^{\rm g}(Q)\,,\label{gKMS}
\eear
with the gluon spectral density $\rho_{\alpha\beta}^{\rm g}(Q)$ that is a hermitian matrix in $(\alpha,\beta)$, one can finally arrive at after some amount of manipulations (see Appendix 2 in Ref.\cite{Jimenez-Alba:2015bia} for the relevant details)
\bear
\Gamma^{\rm single}&=& {g^2\over 2} C^J_2\int {d^4Q\over (2\pi)^4} \left(n_B(q^0)+n_F(p^0+q^0)\right) (2\pi)\delta(p^0+q^0-\sqrt{(\bm p+\bm q)^2+M^2}) \rho^{\rm g}_{\alpha\beta}(Q)\nonumber\\
&\times& {\rm Tr}\left[\gamma^\beta \gamma^0 {\cal P}(\bm p+\bm q) \gamma^\alpha \gamma^0 {\cal P}(\bm p)\right]\,,
\label{damping}
\eear
which is basically a cut of the self-energy where all internal propagators are replaced by their spectral densities. For the bare jet internal line $S_{(0)}(P+Q)$, it imposes simply the on-shell $\delta$ function on the out-going jet state after the scattering, while the spectral density of the internal gluon line encodes the soft t-channel scatterings with hard LLL quarks or hard thermal gluons. A convenient fact for us is that the internal 1-loop momentum $\bm q$ is nothing but the exchanged momentum in these t-channel scattering with the hard particles, so that one can read off the differential scattering rate $d\Gamma^{\rm single}/d^3\bm q$ by simply writing the result as
\be
\Gamma^{\rm single}=\int {d^3\bm q}{d\Gamma^{\rm single}\over d^3\bm q}\,.
\ee

To find the gluon spectral density after re-summing 1-loop gluon self-energy from the LLL quarks, we start from
\be
(G^{ra}(Q))^{-1}=(G^{ra}_{(0)}(Q))^{-1}-\Pi^{ra}(Q)\,,
\ee
where the inverse refers to the Lorentz indices, and $\Pi^{ra}_{\alpha\beta}(Q)$ is the ra-type gluon self-energy at 1-loop
\be
\Pi^{ra}_{\alpha\beta}(Q)= (ig)^2 T_R N_F \langle  j^r_\alpha(Q) j^a_\beta(-Q) \rangle \,,
\ee
where $j_\alpha$ is the quark color current after color indices are stripped off, and the quark color traces gives $T_R$ which is $1/2$ for fundamental
 and $N_c$ for adjoint representation, and $N_F$ is the number of light flavors.
In our LLL approximation in massless limit, the above current-current correlation function factorizes into a product of 1+1 dimensional correlation function and the transverse density of the LLL states. The former is then easily computed using the well-known bosonization of 1+1 dimensional fermion into a massless real scalar field. These have been recently computed in Ref.\cite{Fukushima:2015wck} and the result is given by
\be
\Pi^{ra}_{\alpha\beta}(Q)= \chi \left(Q_\parallel^2 \eta_{\parallel\alpha\beta}-Q_{\parallel\alpha}Q_{\parallel\beta}\right)\,, \quad
\chi\equiv -i {g^2\over\pi} T_R N_F \left(eB\over 2\pi\right) e^{-{{\bm q}_\perp^2\over 2eB}} {1\over Q_{\parallel\epsilon}^2}\,,
 \label{pira}
\ee
where $Q_\parallel$ and $\eta_{\parallel\alpha\beta}$ refer to 1+1 dimensional components of momentum and the metric along the magnetic field direction, ${\bm q}_\perp$ is the component perpendicular to the magnetic field direction, and
\be
Q_{\parallel\epsilon}^2
\equiv Q_\parallel^2\Big|_{q^0\to q^0+i\epsilon}=-(q^0+i\epsilon)^2+q_z^2\,.
\ee

The $(G^{ra}_{(0)}(Q))^{-1}$ and therefore $G^{ra}(Q)$ needs a gauge-fixing, and we choose to work in the covariant gauge where
\be
(G^{ra}_{(0)}(Q))^{-1}=i\left(Q^2 \eta_{\alpha\beta}-Q_\alpha Q_\beta+{1\over\xi}Q_\alpha Q_\beta\right)\bigg|_{q^0\to q^0+i\epsilon}\,,
\ee
where $\xi$ is a gauge parameter. Then, $G^{ra}(Q)$ with (\ref{pira}) is found to be given by
\be
G^{ra}_{\alpha\beta}(Q)=  -i{\eta_{\alpha\beta}\over Q^2_\epsilon}+i(1-\xi){Q_\alpha Q_\beta\over (Q^2_\epsilon)^2}
-\left(Q_\parallel^2 \eta_{\parallel\alpha\beta}-Q_{\parallel\alpha}Q_{\parallel\beta}\right){\chi\over Q^2_\epsilon(Q^2_\epsilon+i\chi Q_\parallel^2)}\,,
 \label{gra}
\ee
where $Q^2_\epsilon\equiv -(q^0+i\epsilon)^2+\bm q^2$.
The gluon spectral density is defined to be twice of the hermitian part of $G^{ra}(Q)$, and since the above is symmetric in Lorentz indices, it is simply twice of the real part: $\rho^{\rm g}_{\alpha\beta}(Q)=2 \,{\rm Re}\left[G^{ra}_{\alpha\beta}(Q)\right]$.

The second term involving $\xi$ is proportional to $Q_\alpha$, which vanishes after being contracted with the jet current $\bar U(P+Q)\gamma^\alpha U(P)$ in (\ref{damping}) by Ward identity\footnote{This can be seen from the fact that the projection operator ${\cal P}(\bm p)$ consists of
the spinors, that is, ${\cal P}(\bm p)\sim U(P)\bar U(P)\gamma^0$. See also (\ref{ward}).}, which ensures the gauge invariance of the scattering rate (\ref{damping}).
From the on-shell constraint in (\ref{damping}) the momentum transfer $Q$ is space like, so the real part from the first term in (\ref{gra})
which is $\sim \delta(Q^2){\rm sgn}(q^0)$ does not contribute to the scattering rate in (\ref{damping}).
The contribution from the last term in (\ref{gra}) represents the scatterings with the LLL states we are looking for.
A simple, but careful computation as in Ref.\cite{Fukushima:2015wck} gives
\be
\rho^{\rm g}_{\alpha\beta}(Q)\sim {(2\pi) Q_{\parallel\alpha}Q_{\parallel\beta} {g^2\over\pi} T_R N_F \left(eB\over 2\pi\right)e^{-{{\bm q}_\perp^2\over 2eB}}{\rm sgn}(q^0)\delta(Q_\parallel^2)\over \left({\bm q}_\perp^2+{g^2\over\pi} T_R N_F \left(eB\over 2\pi\right)e^{-{{\bm q}_\perp^2\over 2eB}}\right)^2}\,,
 \label{gluonrho}
\ee
which is a key ingredient in our subsequent computations.

Since
\be
{\rm sgn}(q^0)\delta(Q_\parallel^2)={1\over 2q^0}\left(\delta(q^0-q_z)+\delta(q^0+q_z)\right)\,,\label{df}
\ee
where we assume the magnetic field points to the $\hat z$ direction, there are two separate pieces in the above spectral function.
They reflect the two light-like spectrums of 1+1 dimensional LLL quarks moving in opposite directions, each corresponding to a definite 4D chirality of massless quarks. Since the gluon vertex with the quarks does not mix the two chiralities, the momentum transfer $Q$ should be given by the momentum difference of the two states within the same 1+1 dimensional chiral spectrum, and therefore $Q$ should be also light-like in 1+1 dimensions. The term with $\delta(q^0-q_z)$ arises from the LLL quarks moving to $\hat z$ direction, while the term with $\delta(q^0+q^z)$ corresponds to the LLL quarks moving to the opposite direction.

Computing the spinor trace in (\ref{damping}) gives
\bear
 &&{\rm Tr}\left[\gamma^\beta \gamma^0 {\cal P}(\bm p+\bm q) \gamma^\alpha \gamma^0 {\cal P}(\bm p)\right]\nonumber\\
 &=&\hat v^\alpha_{\bm p}\hat v^\beta_{\bm p+\bm q}+\hat v^\alpha_{\bm p+\bm q}\hat v^\beta_{\bm p}-\eta^{\alpha\beta}{(P\cdot Q)\over E_{\bm p}E_{\bm p+\bm q}}\equiv S^{\alpha\beta}\,,\label{strace}
 \eear
where $E_{\bm p}\equiv \sqrt{\bm p^2+M^2}$ and
\be
\hat v^\alpha_{\bm p}\equiv {P^\alpha\over E_{\bm p}}=(1,\bm p/E_{\bm p})=(1,\bm v_{\bm p})\,,
\ee
where $\bm v_{\bm p}$ is nothing but the velocity of the jet of momentum $P$.
In deriving the above result, we used the on-shell condition $P^2=-M^2$.
From the above expression, it is straightforward to see the on-shell Ward identity that we claimed before holds
\be
S^{\alpha\beta}Q_\alpha={1\over E_{\bm p}E_{\bm p+\bm q}}(2P\cdot Q+Q^2) P^\beta =0\,,\label{ward}
\ee
where we used the fact that
the energy $\delta$ function in (\ref{damping}) imposes the on-shell condition $(P+Q)^2=-M^2$ which is equivalent to
\be
2P\cdot Q+Q^2=0\,,\label{id1}
\ee
since $P^2=-M^2$.

The scattering rate (\ref{damping}) with the gluon spectral density (\ref{gluonrho}) and the spinor trace (\ref{strace}) are the basic ingredients in our computation of jet quenching parameter in weak coupling theory in the following subsections.

\subsection{$\hat q$ when the jet is parallel to the magnetic field}

Let us first consider the case where the jet is moving parallel to the magnetic field, say along $\hat z$ direction: $\bm p=p_z \hat z$, $p_z>0$.
In this case, the notions of $\parallel$ and $\perp$ from the magnetic field and the jet coincide, so we can use them for both.
From the gluon spectral density (\ref{gluonrho}) and
\be
{\rm sgn}(q^0)\delta(Q_\parallel^2)={1\over 2q^0}\left(\delta(q^0-q_z)+\delta(q^0+q_z)\right)\,,
\label{del}
\ee
there are two distinct delta-functions which give different characteristic contributions to the jet scattering rate. We will find that the one coming from LLL quarks moving opposite to the jet direction (i.e. the one with $\delta(q^0+q_z)$) gives the dominant contribution in high energy limit $v\to 1$.

From (\ref{damping}) with (\ref{gluonrho}), we see that we need to compute $S^{\alpha\beta}Q_{\parallel\alpha}Q_{\parallel\beta}$.
Due to the Ward identity and $Q_{\parallel\alpha}=Q_\alpha-Q_{\perp\alpha}$, this is equal to
\be
S^{\alpha\beta}Q_{\parallel\alpha}Q_{\parallel\beta}=S^{\alpha\beta}Q_{\perp\alpha}Q_{\perp\beta}=-{1\over E_{\bm p}E_{\bm p+\bm q}} (P\cdot Q) \bm q_\perp^2={1\over 2 E_{\bm p}E_{\bm p+\bm q}}Q^2 \bm q^2_{\perp}={(\bm q_\perp^2)^2\over 2E_{\bm p}E_{\bm p+\bm q}}\,,\label{sQQ}
\ee
where we used $P\cdot Q_\perp=0$ and (\ref{id1}), as well as $Q^2=\bm q_\perp^2$ in the last equality due to the $\delta(Q_\parallel^2)$ factor in (\ref{gluonrho}). The net result is quite simple.

From (\ref{del}), let us consider each delta-function separately, and perform $q^0$ integral so that we can replace $q^0$ with $\pm q_z$ where
$\pm$ refers to each case of the two delta-functions. Then, the energy delta function in (\ref{damping}) is worked out as
\bear
&&\delta(p^0+q^0-\sqrt{(\bm p+\bm q)^2+M^2})=\delta\left(\sqrt{p_z^2+M^2}\pm q_z-\sqrt{(p_z+q_z)^2+{\bm q}_\perp^2+M^2}\right )\nonumber\\
&=& {E_{\bm p+\bm q}\over E_{\bm p} (1\mp v)}\delta\left(q_z\mp {{\bm q}_\perp^2\over 2E_{\bm p} (1\mp v)}\right)\,,\label{del2}
\eear
where $E_{\bm p+\bm q}$ should be replaced by
\be
E_{\bm p+\bm q}=E_{\bm p}\pm q_z=E_{\bm p}+{{\bm q}_\perp^2\over 2E_{\bm p} (1\mp v)}\,,
\ee
and
\be
q^0=\pm q_z={{\bm q}_\perp^2\over 2E_{\bm p} (1\mp v)}\,.
\ee

Finally, the statistical factor $(n_B(q^0)+n_F(p^0+q^0))$ in (\ref{damping}) is simplified if we assume that the coupling $\alpha_s=g_s^2/(4\pi)$ is small enough that
\be
q^0={{\bm q}_\perp^2\over 2E_{\bm p} (1\mp v)} \ll T\,,\label{cond2}
\ee
since we will see shortly that the
typical momentum transfer is ${\bm q}_\perp^2 \sim \alpha_s eB$. Then we have at leading order
\be
n_B(q^0)\approx {T\over q^0}={2TE_{\bm p}(1\mp v)\over {\bm q}_\perp^2}\,,\label{nB}
\ee
while $n_F(p^0+q^0)$ is exponentially suppressed due to high energy limit $p^0=E_{\bm p}\to\infty$.

Gathering all the above discussions, especially (\ref{sQQ}), (\ref{del2}) and (\ref{nB}), we finally arrive at a compact result for the scattering rate (\ref{damping}) as
\be
\Gamma^{\rm single}=\sum_{\pm}(8\pi)\alpha_s C_2^J  (1\mp v) T\int {d^2 {\bm q}_\perp\over (2\pi)^2}
{\alpha_s T_RN_F \left(eB\over 2\pi\right) e^{-{{\bm q}_\perp^2\over 2eB}}\over \left({\bm q}_\perp^2+4\alpha_sT_RN_F
\left(eB\over 2\pi\right) e^{-{{\bm q}_\perp^2\over 2eB}}\right)^2}\,,\label{damp2}
\ee
from which we see that the lower sign case (that is, from $\delta(q^0+q_z)$ piece in the gluon spectral density coming from the LLL quarks moving opposite to the jet direction) gives the dominant contribution in high energy limit $v\to 1$.

The condition (\ref{cond2}) we assumed is perfectly fine for the lower sign case (that is, $(1+v)$, or $\delta(q^0+q_z)$ case) in high energy limit: $v\to 1$ and $E_{\bm p}=M\gamma\to \infty$. For the uppers sign case, (\ref{cond2}) will eventually be violated in ultra-high energy limit when
\be
\gamma(1-v)\sim \sqrt{1-v}\lesssim {{\bm q}_\perp^2\over TM} \approx {\alpha_s eB\over TM}\,,
\ee
but in this case, $n_B(q^0)\sim e^{-q^0/T}\ll 1$ is exponentially suppressed anyway. Therefore, we always get the dominant contribution from the $\delta(q^0+q_z)$ piece in the gluon spectral density in high energy limit $v\to 1$, while $\delta(q^0-q_z)$ contribution is sub-leading. We will keep only the dominant contribution in the following.

From (\ref{damp2}), we get the sought-for differential scattering rate of the jet with the LLL quarks
\be
{d\Gamma^{\rm single}\over d^2 {\bm q}_\perp}=
{2\over\pi}\alpha_s C_2^J  (1+v) T
{\alpha_s T_RN_F \left(eB\over 2\pi\right) e^{-{{\bm q}_\perp^2\over 2eB}}\over \left({\bm q}_\perp^2+4\alpha_sT_RN_F
\left(eB\over 2\pi\right) e^{-{{\bm q}_\perp^2\over 2eB}}\right)^2}\,,
\ee
and the jet quenching parameter to complete leading order in $\alpha_s$ is finally computed as
\be
\hat q\equiv{1\over v}\int d^2 {\bm q}_\perp\,{d\Gamma^{\rm single}\over d^2 {\bm q}_\perp} {\bm q}_\perp^2={1\over\pi}(1+1/v)C_2^JT_RN_F\,\alpha_s^2  (eB) T\Big(\log\left(1/\alpha_s\right)-1-\gamma_E-\log\left(T_RN_F/\pi\right)\Big)\,,
\ee
where $\gamma_E\approx 0.577 $ and the leading logarithm is produced from the range
\be
\alpha_s eB \ll {\bm q}_\perp^2 \ll eB\,.
\ee
In getting the above complete leading order result (leading log and the constant under the log), we used the standard technique  \cite{Braaten:1991dd} of
introducing the intermediate scale $\sqrt{\alpha_s eB}\ll q^* \ll \sqrt{eB}$, and divide the integral into two separate regions $|\bm q_\perp|<q^*$ and $|\bm q_\perp|>q^*$ where the integrand simplifies to leading order in $q^*/\sqrt{eB}$ and $\sqrt{\alpha_s eB}/q^*$ (see the next section for a more detailed example of the same technique).
It is interesting to point out that the UV cut-off is provided by the inverse size of the LLL levels, $\sqrt{eB}$, from the exponential term $e^{-{{\bm q}_\perp^2\over 2eB}}$, which is naturally expected
since the LLL states cannot provide or absorb transverse momentum greater than this. It should be also remarked that the jet-quenching parameter from the LLL states is finite in the infinite energy limit of $v\to 1$.

\subsection{$\hat q$ when the jet is perpendicular to the magnetic field}

Let us next consider the case where the jet is moving perpendicular to the magnetic field direction.
We choose the magnetic field to point to $\hat z$, and the jet to move to $\hat x$ direction: $\bm p=p_x\hat x$.
What we mean by ${\bm q}_\perp$ in the gluon spectral density (\ref{gluonrho}) is then ${\bm q}_\perp=(q_x,q_y)$, while the parallel component is $Q_{\parallel}=(q^0,q_z)$. The transverse directions to the jet is $(q_y,q_z)$, and recall that $\hat q$ is defined as a  momentum diffusion constant in this transverse space.

The definition of $\hat q$ assumes a rotational symmetry around the jet direction $\hat x$, which is clearly broken by the magnetic field along $\hat z$. This means that the transverse momentum diffusion of the jet along $\hat z$ will in general be different from the diffusion along $\hat y$ direction. Let us denote the momentum diffusion along $\hat z$ as $\hat q_z$, and along $\hat y$ as $\hat q_y$. The original definition of $\hat q$ assuming the rotational invariance is the sum of momentum diffusion constants along the two transverse directions: $\hat q=\hat q_z+\hat q_y$. The asymmetry in the momentum diffusion constants should affect the BDMPS-Z gluon Bremstrahlung emission pattern in interesting ways to have an azimuthal asymmetry in the gluon emission spectrum (see our discussion in the section \ref{summary}).

From (\ref{damping}) with (\ref{gluonrho}) and (\ref{strace}), we need to compute $S^{\alpha\beta}Q_{\parallel\alpha}Q_{\parallel\beta}=S^{\alpha\beta}Q_{\perp\alpha}Q_{\perp\beta}$ where we again used the Ward identity.
We have
\bear
S^{\alpha\beta}Q_{\perp\alpha}Q_{\perp\beta}&=&{1\over E_{\bm p}E_{\bm p+\bm q}}\left(2(P\cdot Q_\perp)((P+Q)\cdot Q_{\perp})-(P\cdot Q)Q_\perp^2\right)\nonumber\\
&=&{1\over E_{\bm p}E_{\bm p+\bm q}}\left(2 (p_xq_x)(p_xq_x+q_x^2+q_y^2)+{1\over 2}\left(q_x^2+q_y^2\right)^2\right)\,,\label{sQQ2}
\eear
where we used the on-shell condition $2P\cdot Q+Q^2=0$ as well as $Q_\parallel^2=0$ from (\ref{gluonrho}).
We will consider a high jet energy limit such that
\be
p_x\sim M\gamma \gg \sqrt{eB}\gg T\,,
\ee
and since we will see later that $Q\lesssim \sqrt{eB}$, this means that the jet energy is much larger than the momentum transfer: $p_x\sim E_{\bm p}\gg Q$. Then (\ref{sQQ2}) is simplified as
\be
S^{\alpha\beta}Q_{\perp\alpha}Q_{\perp\beta}\approx 2q_x^2{p_x^2\over E_{\bm p}^2}=2q_x^2 v^2\,,\label{sQQ3}
\ee
where $v=p_x/E_{\bm p}$ is the velocity of the jet.

As before, the gluon spectral density (\ref{gluonrho}) has two separate pieces, each from $\delta(q^0\mp q_z)$ (see (\ref{df})).
Performing $q^0$ integration simply replaces $q^0$ with $\pm q_z$. Then the energy $\delta$-function in (\ref{damping})
becomes after some algebra
\bear
&&\delta\left(p^0+q^0-\sqrt{(\bm p+\bm q)^2+M^2}\right)=\delta\left(\sqrt{p_x^2+M^2}\pm q_z-\sqrt{(p_x+q_x)^2+q_y^2+q_z^2+M^2}\right)
\nonumber\\
&=&{(E_{\bm p}\pm q_z)\over \sqrt{p_x^2\pm 2q_z E_{\bm p}-q_y^2}}\left(\delta(q_x+p_x-\sqrt{p_x^2\pm 2 q_z E_{\bm p}-q_y^2})+
\delta(q_x+p_x+\sqrt{p_x^2\pm 2 q_z E_{\bm p}-q_y^2})\right)\nonumber\\
&\sim &{(E_{\bm p}\pm q_z)\over \sqrt{p_x^2\pm 2q_z E_{\bm p}-q_y^2}}\,\,\delta(q_x+p_x-\sqrt{p_x^2\pm 2 q_z E_{\bm p}-q_y^2})\,,\label{d2}
\eear
where in the final form, we dropped the second $\delta$-function, since it would give no contribution due to $Q\ll p_x$.
On the other hand, the first $\delta$-function will put $q_x$ to be
\be
q_x=\sqrt{p_x^2\pm 2 q_z E_{\bm p}-q_y^2}-p_x={\pm 2 q_z E_{\bm p}-q_y^2\over \sqrt{p_x^2\pm 2 q_z E_{\bm p}-q_y^2}+p_x}
\approx \pm q_z {E_{\bm p}\over p_x}=\pm {q_z\over v}\,,
\ee
where we used $p_x\gg Q$ as before. Since $q_x$ is along the jet direction, while we are interested in computing the transverse momentum diffusion along $\hat z$ and $\hat y$ ($\hat q_z$ and $\hat q_y$), we should integrate over $q_x$ at this stage, and the above energy $\delta$-function simply replaces $q_x$ with $\pm q_z/v$ at leading order. The Jacobian in front of the $\delta$-function (\ref{d2}) also simplifies as
\be
{(E_{\bm p}\pm q_z)\over \sqrt{p_x^2\pm 2q_z E_{\bm p}-q_y^2}}\approx {E_{\bm p}\over p_x}={1\over v}\,.
\ee
With all these, the (\ref{sQQ3}) becomes
\be
S^{\alpha\beta}Q_{\perp\alpha}Q_{\perp\beta}\approx 2q_x^2 v^2\approx 2q_z^2\,,
\ee
and the jet scattering rate is given by
\be
\Gamma^{\rm single}\approx \sum_{\pm}{2\over\pi v}\alpha_s C_2^J\int dq_z\int dq_y n_B(\pm q_z)(\pm q_z){\alpha_s T_R N_F \left(eB\over 2\pi\right)e^{-{\left({q_z^2/ v^2}+q_y^2\right)\over 2eB}}\over \left({q_z^2\over v^2}+q_y^2+4\alpha_s T_RN_F
\left(eB\over 2\pi\right)e^{-{\left({q_z^2/ v^2}+q_y^2\right)\over 2eB}}\right)^2}\,.
\ee
For the lower sign (that is coming from $\delta(q^0+q_z)$ piece in the gluon spectral density), we can simply change the variable from $q_z$ to $-q_z$ to get the same expression to the upper sign case, which means that the LLL states moving along or opposite directions to the magnetic field give the same contributions to the jet scattering rate and hence to the momentum diffusion constants.
Therefore, the total scattering rate should be twice of the one with the upper sign and the differential scattering rate we can use in order
to compute the momentum diffusion constants is finally given as
\be
{d\Gamma^{\rm single}\over dq_y dq_z}\approx {4\over\pi v}\alpha_s C_2^J\,n_B(q_z)\,q_z{\alpha_s T_R N_F \left(eB\over 2\pi\right)e^{-{\left({q_z^2/ v^2}+q_y^2\right)\over 2eB}}\over \left({q_z^2\over v^2}+q_y^2+4\alpha_s T_RN_F
\left(eB\over 2\pi\right)e^{-{\left({q_z^2/ v^2}+q_y^2\right)\over 2eB}}\right)^2}\,,\label{gqq}
\ee
which is our starting point of computing the jet quenching parameters $\hat q_z$ and $\hat q_y$ in high energy limit:
\be
\hat q_z={1\over v}\int dq_y\int dq_z\,q_z^2 \,{d\Gamma^{\rm single}\over dq_y dq_z}\,,\quad \hat q_y={1\over v}\int dq_y\int dq_z\,q_y^2 \,{d\Gamma^{\rm single}\over dq_y dq_z}\,.
\ee

One aspect of the above result (\ref{gqq}) is that it contains the vacuum contribution which can be obtained in $T\to 0$ limit.
In $T\to 0$ limit, we have
\be
n_B(q_z) \to -\Theta(-q_z)\,,\quad T\to 0\,,
\ee
which restricts the integral to $q^0=q_z<0$ region. The $q^0<0$ means that the jet {\it gives} the energy to the LLL states, and it is not difficult to find that the only way this is possible in the vacuum is a pair-creation of quark and antiquark pair from the vacuum.
In the presence of the magnetic field with the 1+1 dimensional dispersion relation of LLL quarks, this pair-creation by the jet energy transfer to LLL states
is consistent with the on-shell kinematics, which gives a finite contribution to the jet scattering rate even in the vacuum, as is given by (\ref{gqq}) with $n_B(q_z)\to -\Theta(-q_z)$.

We first compute these vacuum contributions to $\hat q_z$ and $\hat q_y$. We show some details for $\hat q_z^{\rm vacuum}$ and the computation for $\hat q_y^{\rm vacuum}$ is nearly identical.
We have
\bear
\hat q_z^{\rm vacuum}={4\over\pi v^2}\alpha_s C_2^J \int_{-\infty}^\infty dq_y \int_{-\infty}^0 dq_z\,\, (-q_z)^3
{\alpha_s T_R N_F \left(eB\over 2\pi\right)e^{-{\left({q_z^2/ v^2}+q_y^2\right)\over 2eB}}\over \left({q_z^2\over v^2}+q_y^2+4\alpha_s T_RN_F
\left(eB\over 2\pi\right)e^{-{\left({q_z^2/ v^2}+q_y^2\right)\over 2eB}}\right)^2}\,.\nonumber\\
\eear
Changing $q_z\to v q_z$, and working in the polar coordinate system of $(q_z,q_y)$ plane, $(q,\theta)$, we have
\bear
\hat q_z^{\rm vacuum}={4v^2\over\pi}\alpha_s C_2^J\int_{\pi/2}^{3\pi/2} d\theta(-\cos\theta)^3\int_0^\infty dq\,\, q^4 {\alpha_s T_R N_F \left(eB\over 2\pi\right)e^{-{q^2\over 2eB}}\over \left(q^2+4\alpha_s T_RN_F
\left(eB\over 2\pi\right)e^{-{q^2\over 2eB}}\right)^2}\,.
\eear
Without the exponential factor in the numerator, the $q$ integral is linearly divergent in large $q$ limit, so the exponential factor in the numerator provides a relevant UV cutoff, which implies that the dominant leading contribution to the final result comes from the region $q^2\sim eB$.
Then in the denominator, one can safely neglect the Debye mass term which is $m^2_{D,B}\sim\alpha_s eB\ll eB\sim q^2$ compared to $q^2$ at leading order computation. This brings us to leading order
\bear
\hat q_z^{\rm vacuum}&=&{4v^2\over\pi}\alpha_s C_2^J\int_{\pi/2}^{3\pi/2} d\theta(-\cos\theta)^3\int_0^\infty dq\,\,\alpha_s T_R N_F \left(eB\over 2\pi\right)e^{-{q^2\over 2eB}}\nonumber\\
&=&{16 v^2\over 3(2\pi)^{3/2}} C_2^J T_RN_F \alpha_s^2 (eB)^{3/2}\,.\label{qzvac}
\eear
The next-to-leading order correction is further suppressed by an additional factor of $\sqrt{\alpha_s}$ coming from the region $q\sim \sqrt{\alpha_s eB}$.
The almost same computation gives the leading order vacuum contribution to $\hat q_y$ as
\bear
\hat q_y^{\rm vacuum}&=&{4\over\pi}\alpha_s C_2^J\int_{\pi/2}^{3\pi/2} d\theta(-\cos\theta\sin^2\theta)\int_0^\infty dq\,\,\alpha_s T_R N_F \left(eB\over 2\pi\right)e^{-{q^2\over 2eB}}\nonumber\\
&=&{8 \over 3(2\pi)^{3/2}} C_2^J T_RN_F \alpha_s^2 (eB)^{3/2}\,.
\eear
We see that $\hat q_z^{\rm vacuum}\neq \hat q_y^{\rm vacuum}$ at leading order, which implies that the momentum diffusion in the transverse space of the jet direction is asymmetric.

Next, we would like to compute the thermal contributions at finite temperature $T$. This can be obtained by subtracting the vacuum contribution from (\ref{gqq}):
\be
{d\Gamma^{\rm single}_{\rm thermal}\over dq_y dq_z}\approx {4\over\pi v}\alpha_s C_2^J\,\left(n_B(q_z)+\Theta(-q_z)\right)\,q_z{\alpha_s T_R N_F \left(eB\over 2\pi\right)e^{-{\left({q_z^2/ v^2}+q_y^2\right)\over 2eB}}\over \left({q_z^2\over v^2}+q_y^2+4\alpha_s T_RN_F
\left(eB\over 2\pi\right)e^{-{\left({q_z^2/ v^2}+q_y^2\right)\over 2eB}}\right)^2}\,.\label{thermal}
\ee
From the fact that
\be
n_B(q_z)+\Theta(-q_z)\approx {\rm sgn}(q_z)e^{-|q_z|/T}\,,\quad |q_z|\gg T\,,
\ee
the integration range of $q_z$ is effectively confined into $|q_z|\lesssim T$. Then, due to the hierarchy we are assuming $eB\gg T^2$, we can replace the exponent $e^{-{q_z^2/v^2\over 2eB}}$ with $1$ at leading order in $T^2/eB\ll 1$:
\be
{d\Gamma^{\rm single}_{\rm thermal}\over dq_y dq_z}\approx {4\over\pi v}\alpha_s C_2^J\,\left(n_B(q_z)+\Theta(-q_z)\right)\,q_z{\alpha_s T_R N_F \left(eB\over 2\pi\right)e^{-{q_y^2\over 2eB}}\over \left({q_z^2\over v^2}+q_y^2+4\alpha_s T_RN_F
\left(eB\over 2\pi\right)e^{-{q_y^2\over 2eB}}\right)^2}\,.\label{thermal}
\ee

There are three important scales in the above result: 1) $\sqrt{\alpha_s eB}$ which sets the scale of Debye screening mass (that appears in the denominator) which serves an IR cut-off, 2) the temperature $T$ that enters $n_B(q_z)+\Theta(-q_z)$, 3) $\sqrt{eB}$ that gives the ultimate UV cutoff by the exponential suppression $e^{-{q_y^2\over 2eB}}$. Recall that our assumption on hierarchy of scales is $\sqrt{\alpha_s eB}\ll T\ll\sqrt{eB}$. It can be easily seen from the $q_y$ integral in (\ref{thermal}) that the leading contribution
comes from the region
\be
|q_y|\sim \sqrt{q_z^2/v^2+\alpha_s eB}\lesssim T\,.
\ee
This is because $q_y$ integral is UV convergent for both $\hat q_z$ and $\hat q_y$ due to the denominator, independent of the existence
of the $e^{-{q_y^2\over 2eB}}$ term. Therefore, to leading order in $T^2/eB$ we again can replace $e^{-{q_y^2\over 2eB}}$ with 1,
and we finally have
\be
{d\Gamma^{\rm single}_{\rm thermal}\over dq_y dq_z}\approx {4\over\pi v}\alpha_s C_2^J\,\left(n_B(q_z)+\Theta(-q_z)\right)\,q_z{\alpha_s T_R N_F \left(eB\over 2\pi\right)\over \left({q_z^2\over v^2}+q_y^2+4\alpha_s T_RN_F
\left(eB\over 2\pi\right)\right)^2}\,,\label{thermal2}
\ee
valid at leading order. This means that the ultimate UV cutoff, $\sqrt{eB}$, does not play a role at leading order in $T^2/eB$, and the leading order result comes from the softer scale dynamics between $\sqrt{\alpha_s eB}$ and $T$.

Let us show some details of our computation of $\hat q_z$ with (\ref{thermal2}) at complete leading order in $\alpha_s$ (that is, the leading log as well as the constant under the log):
\bear
\hat q_z^{\rm thermal}&\equiv&{4\over\pi v^2}\alpha_s C_2^J\int dq_z\int dq_y \,\left(n_B(q_z)+\Theta(-q_z)\right)\,q_z^3{\alpha_s T_R N_F \left(eB\over 2\pi\right)\over \left({q_z^2\over v^2}+q_y^2+4\alpha_s T_RN_F
\left(eB\over 2\pi\right)\right)^2}\nonumber\\
&=&{2\over v^2}\alpha_s^2 C_2^JT_R N_F \left(eB\over 2\pi\right)\int dq_z \,\left(n_B(q_z)+\Theta(-q_z)\right)\,q_z^3{1\over \left({q_z^2\over v^2}+4\alpha_s T_RN_F
\left(eB\over 2\pi\right)\right)^{3\over 2}}\,,\nonumber\\\label{qz}
\eear
where we performed the $q_y$ integration in the last line. It is not difficult to see from the above that the remaining $q_z$ integral produces the logarithm between the IR cutoff $\sqrt{\alpha_s eB}$ and the UV cutoff $T$. To handle this,
we follow the standard technique \cite{Braaten:1991dd} of introducing an intermediate scale $q^*$ between $\sqrt{\alpha_s eB}$ and $T$ (that is, $\sqrt{\alpha_s eB}\ll q^*\ll T$), and divide the $q_z$ integral into $|q_z|<q^*$ and $|q_z|>q^*$.
In the first integral of $|q_z|<q^*$, since $|q_z|\ll T$ we can replace to leading order
\be
n_B(q_z)+\Theta(-q_z)\approx {T\over q_z}\,,
\ee
and we have
\bear
&&{2\over v^2}\alpha_s^2 C_2^JT_R N_F \left(eB\over 2\pi\right)T\int_{-q^*}^{q^*} dq_z \,q_z^2{1\over \left({q_z^2\over v^2}+4\alpha_s T_RN_F
\left(eB\over 2\pi\right)\right)^{3\over 2}}\nonumber\\ &=&
{2v}\alpha_s^2 C_2^JT_R N_F \left(eB\over 2\pi\right)T\left(\log\left((q^*)^2\over \alpha_s T_RN_F\left(eB\over 2\pi\right) v^2\right)-2+{\cal O}\left(\alpha_s eB\over (q^*)^2\right)\right)\,.\label{qz1}
\eear
In the other region of $|q_z|>q^*$, we instead have $|q_z|\gg\sqrt{\alpha_s eB}$, so we can ignore the Debye mass in the denominator at leading order to have
\bear
&&{2v}\alpha_s^2 C_2^JT_R N_F \left(eB\over 2\pi\right)\int_{|q_z|>q^*} dq_z \,\left(n_B(q_z)+\Theta(-q_z)\right){\rm sgn}(q_z)\nonumber\\&=&{2v}\alpha_s^2 C_2^JT_R N_F \left(eB\over 2\pi\right)T\left(\log\left(T^2\over (q^*)^2\right)+{\cal O}\left(q^*\over T\right)\right)\,.\label{qz2}
\eear
Combining the two regions (\ref{qz1}) and (\ref{qz2}), we finally have the thermal contribution to $\hat q_z^{\rm thermal}$ at complete leading order as
\be
\hat q_z^{\rm thermal}={1\over\pi}{v} C_2^JT_R N_F \alpha_s^2(eB)T\left(\log\left(T^2\over \alpha_s T_RN_F\left(eB\over 2\pi\right) v^2\right)-2\right)\,,\label{qzthermfin}
\ee
to leading order in $\alpha_s$ and $\alpha_s eB/T^2$.
Recall our assumed hierarchy of scales $\alpha_s eB\ll T^2\ll eB$.

A similar computation can be done for $\hat q_y^{\rm thermal}$:
\bear
\hat q_y^{\rm thermal}&\equiv&{4\over\pi v^2}\alpha_s C_2^J\int dq_z\int dq_y \,\left(n_B(q_z)+\Theta(-q_z)\right)\,q_z q_y^2{\alpha_s T_R N_F \left(eB\over 2\pi\right)\over \left({q_z^2\over v^2}+q_y^2+4\alpha_s T_RN_F
\left(eB\over 2\pi\right)\right)^2}\nonumber\\
&=&{2\over v^2}\alpha_s^2 C_2^JT_R N_F \left(eB\over 2\pi\right)\int dq_z \,\left(n_B(q_z)+\Theta(-q_z)\right)\,q_z{1\over \left({q_z^2\over v^2}+4\alpha_s T_RN_F
\left(eB\over 2\pi\right)\right)^{1\over 2}}\,.\nonumber\\\label{qy}
\eear
 From the region $|q_z|<q^*$ we have
\bear
&&{2\over v^2}\alpha_s^2 C_2^JT_R N_F \left(eB\over 2\pi\right)T\int_{-q^*}^{q^*} dq_z \,{1\over \left({q_z^2\over v^2}+4\alpha_s T_RN_F
\left(eB\over 2\pi\right)\right)^{1\over 2}}\nonumber\\ &=&
{2\over v}\alpha_s^2 C_2^JT_R N_F \left(eB\over 2\pi\right)T\left(\log\left((q^*)^2\over \alpha_s T_RN_F\left(eB\over 2\pi\right) v^2\right)+{\cal O}\left(\alpha_s eB\over (q^*)^2\right)\right)\,.\label{qy1}
\eear
and from the region $|q_z|>q^*$ we have
\bear
&&{2\over v}\alpha_s^2 C_2^JT_R N_F \left(eB\over 2\pi\right)\int_{|q_z|>q^*} dq_z \,\left(n_B(q_z)+\Theta(-q_z)\right){\rm sgn}(q_z)\nonumber\\&=&{2\over v}\alpha_s^2 C_2^JT_R N_F \left(eB\over 2\pi\right)T\left(\log\left(T^2\over (q^*)^2\right)+{\cal O}\left(q^*\over T\right)\right)\,.\label{qy2}
\eear
so the final result for $\hat q_y^{\rm thermal}$ at complete leading order is given by
\be
\hat q_y^{\rm thermal}={1\over\pi v} C_2^JT_R N_F \alpha_s^2(eB)T\left(\log\left(T^2\over \alpha_s T_RN_F\left(eB\over 2\pi\right) v^2\right)+0\right)\,,\label{qythermfin}
\ee
where by $0$ in the above, we mean there is no other constant under the log than what is shown in the above result.

Comparing (\ref{qzthermfin}) and (\ref{qythermfin}), we see that $\hat q_z^{\rm thermal}$ and $\hat q_y^{\rm thermal}$ are in general different, but in the high energy limit $v\to 1$, they differ
only by a constant under the log, while they become equal at leading log order in $T^2/(\alpha_s eB)$.

In summary, the sum of the vacuum and thermal contributions to the $\hat q_z$ and $\hat q_y$ is given by
\bear
\hat q_z&=&{16 v^2\over 3(2\pi)^{3/2}} C_2^J T_RN_F \alpha_s^2 (eB)^{3/2}+{1\over\pi}{v} C_2^JT_R N_F \alpha_s^2(eB)T\left(\log\left(T^2\over \alpha_s T_RN_F\left(eB\over 2\pi\right) v^2\right)-2\right)\,,\nonumber\\
\hat q_y&=&{8 \over 3(2\pi)^{3/2}} C_2^J T_RN_F \alpha_s^2 (eB)^{3/2}+{1\over\pi v} C_2^JT_R N_F \alpha_s^2(eB)T\left(\log\left(T^2\over \alpha_s T_RN_F\left(eB\over 2\pi\right) v^2\right)+0\right)\,.\nonumber\\\label{weaksum}
\eear
We should note that the next-to-leading order correction to the vacuum contribution (the first term in the above) is further suppressed by $\sqrt{\alpha_s}$ compared to the leading order (see the previous discussion below (\ref{qzvac})),
so it is sub-leading by $\sqrt{\alpha_s eB}/T\ll 1$ compared to the leading order result from the thermal contributions (the second term in the above).
Therefore, the above two terms indeed represent the first two leading terms in our assumed hierarchy of scales $\alpha_s eB\ll T^2\ll eB$.

\section{Jet quenching in strongly coupled regime }

In this section, we compute our jet quenching parameter in strong magnetic field in the AdS/CFT correspondence.
We use two well-established methods in literature corresponding to the two different definitions of the jet quenching parameter, albeit the fact that these two definitions agree with each other at weak coupling regime: 1) the first definition is what we have used in our computation at weak coupling, that is, the transverse momentum diffusion constant, $\hat q={d\langle \bm p_\perp^2\rangle\over dz}$, 2) the second definition is in terms of a light-like Wilson loop \cite{Salgado:2003gb} with a transverse spatial separation $\bm b_\perp$ in small $\bm b_\perp$ limit behaving as $\langle W(\bm b_\perp)^\dagger W(\bm 0)\rangle\sim \exp[{-{1\over 4\sqrt{2}}\hat q \bm b_\perp^2 x^+}]$ where $x^+$ is the light-like extension of the loop. To see the equivalence heuristically at weak coupling (we will not be precise about color factors and normalizations), let's prepare a fast moving initial state with a transverse momentum $\bm p_\perp$ written in the position basis $|\bm x_\perp\rangle$ as
\be
|\bm p_\perp\rangle={1\over \sqrt{S_\perp}}\int d^2 \bm x_\perp e^{i\bm p_\perp\cdot\bm x_\perp}|\bm x_\perp\rangle\,,
\ee
where $S_\perp$ is the transverse area put to normalize the state. After traversing the light-like distance $x^+$, each state $|\bm x_\perp\rangle$ in the eikonal approximation will pick-up the Wilson line $W(\bm x_\perp)$, so the final state becomes
\be
|\psi_f\rangle={1\over \sqrt{S_\perp}}\int d^2 x e^{i\bm p_\perp\cdot\bm x_\perp} W(\bm x_\perp)|\bm x_\perp\rangle\,,
\ee
and the transition S-matrix to the state with additional momentum kick $\bm q_\perp$ is
\be
\langle\bm p_\perp+\bm q_\perp|\psi_f\rangle={1\over S_\perp}\int d^2 \bm x_\perp e^{-i\bm q_\perp\cdot\bm x_\perp} W(\bm x_\perp)\,.
\ee
Then, the probability distribution of transverse momentum $P(\bm q_\perp)$ after traversing the light distance $x^+$ becomes
\be
P(\bm q_\perp,x^+)=|\langle\bm p_\perp+\bm q_\perp|\psi_f\rangle|^2={1\over S_\perp}\int d^2 \bm b_\perp e^{i\bm q_\perp\cdot \bm b_\perp}
\langle W(\bm b_\perp)^\dagger W(\bm 0)\rangle\,,
\ee
where we have used the translational invariance in the transverse space.
If the Wilson loop behaves as $\langle W(\bm b_\perp)^\dagger W(\bm 0)\rangle\sim \exp[-{1\over 4\sqrt{2}}\hat q \bm b_\perp^2 x^+]$, the distribution evolves in time (or space $z$) as
\bear
{\partial P(\bm q_\perp,x^+)\over \partial z}&=&\sqrt{2}{\partial P(\bm q_\perp,x^+)\over \partial x^+}={\hat q\over 4}{1\over S_\perp}\int d^2 \bm b_\perp (- \bm b_\perp^2)e^{i\bm q_\perp\cdot \bm b_\perp}
\langle W(\bm b_\perp)^\dagger W(\bm 0)\rangle\nonumber\\&=&{\hat q\over 4}{1\over S_\perp}\int d^2 \bm b_\perp (\bm \nabla_{\bm q_\perp}^2 e^{i\bm q_\perp\cdot \bm b_\perp})
\langle W(\bm b_\perp)^\dagger W(\bm 0)\rangle\nonumber\\
&=&{\hat q\over 4} \bm\nabla_{\bm q_\perp}^2{1\over S_\perp}\int d^2 \bm b_\perp e^{i\bm q_\perp\cdot \bm b_\perp}
\langle W(\bm b_\perp)^\dagger W(\bm 0)\rangle={\hat q\over 4}\bm\nabla_{\bm q_\perp}^2 P(\bm q_\perp,x^+)\,,
\eear
which is precisely the Fokker-Planck equation coming from the random momentum kicks with the momentum diffusion constant $\hat q$, showing the equivalence of the two definitions.

In section \ref{sec31}, we compute $\hat q$ via the definition of 1) in the AdS/CFT correspondence using a single string world-sheet moving with a velocity $v$; the method developed in Refs \cite{CasalderreySolana:2007qw,Gubser:2006nz}. The momentum diffusion constant is identified from the low frequency limit of the spectral density of color electric field correlators in real-time Schwinger-Keldysh formalism, quite similar to conductivity for current operators. In operator-field mapping in the AdS/CFT, the color electric field operator maps to the transverse displacement of the string world-sheet.
Since the low frequency limit of spectral density in AdS/CFT correspondence is given solely by event-horizon properties via membrane paradigm \cite{Iqbal:2008by}, we will skip the details already present in literature, and simply apply the known expression to our situation with strong magnetic field. The same universality has also been derived by holographic RG formalism in low frequency limit.

In section \ref{sec32}, we compute $\hat q$ in the definition of 2) from the light-like Wilson loops; the method used in Ref.\cite{Liu:2006ug,Liu:2006he}. As is the case without magnetic field in literature, the definition 2) gives a different result from that from 1), which still seems to be an open issue.

The black-hole geometry in AdS space with a magnetic field in $z$ direction takes a form
\bear \label{BH}
ds^{2}=g_{zz}\big(-f(r)dt^{2}+dz^{2}\big)+g_{xx}\big(dx^{2}+dy^{2}\big)+\frac{1}{p(r)}dr^{2}\,.
\eear
The Hawking temperature $T$ of the black hole which is identified with the field theory temperature is 
\be \label{T}
T=\frac{1}{4\pi}\sqrt{g_{zz}(r_{h})f'(r_{h})p'(r_{h})}\,,
\ee
where $r_{h}$ is the radius of the black hole horizon which solves $f(r_{h})=0$.
In the presence of a strong magnetic field $B\gg T^2$ in the bulk,  the black hole metric (\ref{BH}) takes the particular form for the region $r\ll \sqrt{{\cal B}}{\cal R}^2$ where the scale is much smaller than the magnetic field \cite{D'Hoker:2009mm}
\be \label{BTZ}
ds^{2}=\frac{r^2}{\mathcal{R}^2}\left(-f(r)dt^2+dz^2\right)+\mathcal{R}^2\mathcal{B}(dx^2+dy^2)+\frac{1}{\frac{r^2}{\mathcal{R}^2}f(r)}dr^2\,,
\ee
where $f(r)=1-\frac{r_{h}^2}{r^2}$ with the horizon corresponding to $r=r_{h}$, and $R^{4}=\lambda\alpha'^{2}$ is the radius of the $AdS_{5}$ spacetime ($\lambda=g_{YM}^2 N_c$ is the strong coupling constant and $\alpha'=l_s^2$ is the string length scale which disappears in final physical results).
The above metric is a product of 3 dimensional BTZ and trivial flat two dimensions. We identify $\mathcal{R}=\frac{R}{\sqrt{3}}$ as the radius of the $AdS_{3}$ spacetime or BTZ black hole, and $\mathcal{B}=\sqrt{3}B=\sqrt{3}F_{xy}$ as the physical magnetic field at the boundary. The Hawking temperature $T$ of the BTZ black hole (\ref{BTZ}) is
\be \label{TBTZ}
T=\frac{1}{4\pi}\sqrt{g_{zz}(r_{h})f'(r_{h})p'(r_{h})}=\frac{r_{h}}{2\pi\mathcal{R}^2}.
\ee

\subsection{$\hat q$ from transverse momentum diffusion\label{sec31}}

The transverse momentum diffusion constant $\kappa(v)$ ``per unit time'' of a heavy quark moving with velocity $v$ in the strongly coupled regime at zero magnetic field, was first computed in Refs.\cite{CasalderreySolana:2007qw, Gubser:2006nz} for ${\cal N}=4$ Super Yang-Mills theory, and was generalized to non-conformal theories in Ref.\cite{Gursoy:2010aa}. In the eikonal regime of high jet energy, there should be no distinction between heavy-quark and the jet for the momentum diffusion constant, since the scatterings would care only about its color charges. Based on this premise, we can identify
\be
\hat q(v)={2\over v}\kappa(v)\,,\label{qfromk}
\ee
where the factor $2$ is from the definition of $\kappa(v)$: it is defined by $\langle \xi_T^i(t)\xi^j_T(t')\rangle=\kappa \delta^{ij}\delta(t-t')$, so that
\be
\kappa={1\over 2}\int d^2\bm q_\perp^2 {d\Gamma\over d^2\bm q_\perp^2} \bm q_\perp^2\,,
\ee
and $1/v$ is from translating $d/dz=(1/v)d/dt$.

\subsubsection{$\hat q$ when the jet is parallel to the magnetic field }

In the presence of strong magnetic field parallel to the jet, the Nambu-Goto (NG) action is
\be \label{NGpar}
S_{NG}^\parallel=\int d\tau d\sigma \mathcal{L}^\parallel(\bar{h}_{ab})=-\frac{1}{2\pi\alpha'}\int d\tau d\sigma \sqrt{-det\,\overline{h}_{ab}}\,,
\ee
where the background induced metric on the string $\overline{h}_{ab}$ is given by
\be \label{ind}
\overline{h}_{ab}=g_{\mu\nu}\partial_{a}x^{\mu}(\tau,\sigma)\partial_{b}x^{\nu}(\tau,\sigma)\,.
\ee

Using the embedding $(\tau, \sigma)\Rightarrow (t(\tau, \sigma),0,0,z(\tau, \sigma), r=\sigma)$, the background induced metric $\overline{h}_{ab}(\dot{z},z')$ (\ref{ind}) becomes (${\cdot} \equiv d/d\tau, {'} \equiv d/d\sigma$)
\be \label{bgindz1}
\overline{h}_{ab}(\dot{z},z')=g_{tt}\partial_{a}t\partial_{b}t+g_{zz}\partial_{a}z\partial_{b}z+g_{rr}\partial_{a}r\partial_{b}r \,.
\ee
Using a particular Ansatz of the form $t(\tau,\sigma)=\tau+K(\sigma)$ and $z=v\tau+F(\sigma)$, which represents a ``trailing string'' configuration moving with velocity $v$, the background induced metric (\ref{bgindz1}) becomes
\bear \label{bgindz2}
\overline{h}_{\tau\tau}(v,z')&=&g_{tt}+v^2g_{zz}\,,\nonumber\\
\overline{h}_{\sigma\sigma}(v,z')&=&g_{tt}(K')^2+g_{zz}(z')^2+g_{rr}\,,\nonumber\\
\overline{h}_{\tau\sigma}(v,z')&=&g_{tt} K'+g_{zz}z'v\,.
\eear
Finding the equation of motion from the action, we have
\be
\partial_\sigma\left({g_{tt} g_{zz}(z'-vK')\over \sqrt{-det\,\overline{h}_{ab}}}\right)=0\,.
\ee

There exists a gauge freedom of re-parametrizing the world-sheet coordinate $\tau$: $\tau\to \tau+h(\sigma)$ for any function $h(\sigma)$, under which we have the transformation $K(\sigma)\to K(\sigma)+h(\sigma)$ and $z\to z+vh(\sigma)$.
Indeed, the above equation of motion is invariant under this transformation, as it should.
Requiring $\overline{h}_{\tau\sigma}(v,z')=0$ to fix this gauge freedom, we have an additional constraint $K'=-\frac{g_{zz}}{g_{tt}}z'v$, which can be used to diagonalize (\ref{bgindz2}) as
\bear \label{bgindz3}
\overline{h}_{\tau\tau}(v,z')&=&-g_{zz}f\Big(1-\frac{v^2}{f}\Big)\,,\nonumber\\
\overline{h}_{\sigma\sigma}(v,z')&=&g_{zz}\Big(1-\frac{v^2}{f}\Big)(z')^2+g_{rr}\,,\nonumber\\
\overline{h}_{\tau\sigma}(v,z')&=&0\,,
\eear
while the equation of motion in this gauge becomes
\be \label{sol1}
\frac{g_{zz}^2f}{\sqrt{-det\,\bar{h}_{ab}} }\left(1-{v^2\over f}\right)z'={\rm constant}\equiv C_{zz}v\,.
\ee
Using $g_{rr}=\frac{1}{g_{zz}f}$ and
\be
-det\,\bar{h}_{ab}=-\bar{h}_{\tau\tau}(v,z')\bar{h}_{\sigma\sigma}(v,z')=g_{zz}^2f\left(1-{v^2\over f}\right)^2(z')^2-(1-\frac{v^2}{f})\,,
\ee
we find
\be\label{sol2}
(z')^2=\frac{C_{zz}^2v^2}{g^4_{zz}f^2}\frac{1}{\left(1-\frac{v^2}{f}\right)\big(1-\frac{C_{zz}^2v^2}{g^2_{zz}f}\big)}\,.
\ee

Since the factor $(1-\frac{v^2}{f})$ in (\ref{sol2}) vanishes when $f(r_{s})=v^2$, requiring $(z')^2$ to be positive across $r=r_s$, the other factor $(1-\frac{C_{zz}^2v^2}{g^2_{zz}f})$ has to vanish at $r=r_{s}$ as well, which will fix the integration constant $C_{zz}=g_{zz}(r_s)$.
Therefore, (\ref{sol2}) becomes
\be\label{sol3}
(z')^2=\frac{g^2_{zz}(r_{s})}{g^4_{zz}(r)}\frac{v^2}{f^2(r)}\frac{1}{\left(1-\frac{v^2}{f(r)}\right)\big(1-\frac{g^2_{zz}(r_{s})}{g^2_{zz}(r)}\frac{v^2}{f(r)}\big)}\,,
\ee
and using this the metric (\ref{bgindz3}) is finally given by
\bear \label{bgindz5}
\overline{h}_{\tau\tau}(v,z')&=&g_{zz}\big(-f+v^2\big)\,,\nonumber\\
\overline{h}_{\sigma\sigma}(v,z')&=&g_{zz}\Bigg(\frac{1}{g^2_{zz}(r)f(r)-g^2_{zz}(r_{s})v^2}\Bigg)\,,\nonumber\\
\overline{h}_{\tau\sigma}(v,z')&=&0\,,
\eear
which can be interpreted as a metric of a 2-dimensional black hole with a line element $ds_{(2)}^2$ given by
\be \label{2dBH}
ds_{(2)}^{2}=\bar{h}_{\tau\tau}d\tau^2+\bar{h}_{\sigma\sigma}d\sigma^2=g_{zz}(-\tilde{f}(r))d\tau^2+\frac{1}{\tilde{p}(r)}d\sigma^2\,,
\ee
where $\tilde{f}(r)=f-v^2$, $\tilde{p}(r)=[g^2_{zz}(r)f(r)-g^2_{zz}(r_{s})v^2](g_{zz})^{-1}$, and the radius of the horizon $r_{s}$ of the 2-dimensional black hole is found from $\tilde{f}(r_s)=0$ or $f(r_{s})=v^2$, i.e., $r_{s}=\gamma r_{h}$ where $\gamma=\frac{1}{\sqrt{1-v^2}}$.

The Hawking temperature of the 2-dimensional black hole denoted as $T_{s}^{\parallel}$ is still given by (\ref{T}) after replacing $T\rightarrow T_{s}^{\parallel}$, $f(r)\rightarrow\tilde{f}(r)$ and $p(r)\rightarrow\tilde{p}(r)$, i.e.,
\be \label{Tspar}
T_{s}^{\parallel}=\frac{1}{4\pi}\sqrt{g_{zz}(r_{s})\tilde{f}'(r_{s})\tilde{p}'(r_{s})}=\frac{r_{h}}{2\pi\mathcal{R}^2}\sqrt{1+v^2}=T\sqrt{1+v^2}\,,
\ee
where we used $\tilde{p}'(r_{s})=2g'_{zz}(r_{s})v^2+g_{zz}(r_{s})f'(r_{s})$ and $r_{s}=\gamma r_{h}$.

Note that the drag force acting on the heavy quark $F_{drag}^{\parallel}$ is simply given by
\be\label{dragz}
F_{drag}^{\parallel}=\frac{\delta\mathcal{L}}{\delta z'}=-\dfrac{C_{zz}}{2\pi\alpha'}v=-\dfrac{2}{3}\pi\sqrt{\lambda}\gamma^2T^2v\,,
\ee
where we used $C_{zz}=g_{zz}(r_s)$ and $r_{s}=\gamma r_{h}$ to get the last line. This is independent of the magnetic field in our limit ${B}\gg T^2$. This could be interpreted as a superfluid nature of the LLL states in strong magnetic field, as discussed in Ref.\cite{Sadofyev:2015tmb} (see also Refs.\cite{RS,SYee}).

To obtain the transverse momentum diffusion constant from the color electric field correlators, we consider the fluctuations of the dual field, that is, the fluctuations of transverse position of the string, $\delta x$.
The transverse fluctuation $\delta h_{ab}(\delta \dot{x},\delta x')$ around the background induced metric $\overline{h}_{ab}(v,z')$ (\ref{bgindz5}) is given by
\bear \label{flucx1}
\delta h_{\tau\tau}(\delta \dot{x},\delta x')&=&g_{xx}(\delta \dot{x})^2\,,\nonumber\\
\delta h_{\sigma\sigma}(\delta \dot{x},\delta x')&=&g_{xx}(\delta x')^2\,,\nonumber\\
\delta h_{\tau\sigma}(\delta \dot{x},\delta x')&=&g_{xx}(\delta \dot{x}\delta x')^2\,.
\eear
Replacing $\overline{h}_{ab}(v,z')\rightarrow \overline{h}_{ab}(v,z')+\delta h_{ab}(\delta \dot{x},\delta x')$ in $S^\parallel_{NG}$ (\ref{NGpar}), and expanding it to linear order in $\delta h_{ab}(\delta \dot{x},\delta x')$, one finds
\bear \label{NGparlinear1}
S^\parallel_{NG}&=&\int d\tau d\sigma \mathcal{L}^\parallel\big(\overline{h}_{ab}(v,z'),\delta h_{ab}(\delta \dot{x},\delta x')\big)\,,\nonumber\\
                &=&-\frac{1}{4\pi\alpha'}\int d\tau d\sigma g_{xx}\sqrt{-det\,\overline{h}_{ab}(v,z')}\overline{h}^{ab}(v,z')\delta h_{ab}(\delta \dot{x},\delta x')\,,\nonumber\\
                &=&-\frac{1}{2}\int d\tau d\sigma \overline{G}_\parallel^{ab}(v,z')\partial_{a}\delta x(\tau,\sigma)\partial_{b}\delta x(\tau,\sigma)\,,
\eear
where $\overline{G}_\parallel^{ab}(v,z')\equiv\frac{1}{2\pi\alpha'}g_{xx}\sqrt{-det\,\overline{h}_{ab}(v,z')}\overline{h}^{ab}(v,z')$. Note that the indices $a$ and $b$ are raised and lowered using the background induced metric $\overline{h}_{ab}(v,z')$, and $\overline{h}^{ab}(v,z')$ is the inverse of $\overline{h}_{ab}(v,z')$.

Using the conjugate momenta $\Pi^\parallel=\frac{\partial\mathcal{L}^\parallel}{\partial_{\sigma}\delta x}$, defining the retarded Green's function $G_{R}^\parallel\equiv-\frac{\Pi^\parallel}{\delta x}$ as in Ref.\cite{Iqbal:2008by}, and using the equation of motion for $\delta x$ in momentum space derived from the action (\ref{NGparlinear1})
\be\label{EOM2}
\partial_{\sigma}\overline{G}_\parallel^{\sigma\sigma}\partial_{\sigma}\delta x-\omega^2\overline{G}_\parallel^{\tau\tau}\delta x=0\,,
\ee
one can derive the holographic RG flow equation for the retarded Green's function $G_{R}^\parallel$ to be
\be\label{RG}
\partial_{\sigma}G_{R}^\parallel=-\frac{(G_{R}^\parallel)^2}{\overline{G}_\parallel^{\sigma\sigma}}+\omega^2\overline{G}_\parallel^{\tau\tau}.
\ee
Since $\overline{G}_\parallel^{\tau\tau}$ and $\frac{1}{\overline{G}_\parallel^{\sigma\sigma}}$ diverge at the horizon of the 2-dimensional black hole metric, i.e., at $r=r_{s}$, we first note that $G_R^\parallel$ vanishes at $\omega=0$, and we expect $G_R^\parallel\propto \omega$ for small $\omega$ limit. Since the right-hand side is ${\cal O}(\omega^2)$, $G_R^\parallel$ becomes a constant in $\sigma$ in $\omega\to 0$ limit. Demanding the regularity of the right-hand side at the horizon, we find
\bear\label{GR}
G_{R}^\parallel(\omega)&=&\pm\omega\sqrt{\overline{G}_\parallel^{\tau\tau}\overline{G}_\parallel^{\sigma\sigma}}\mid_{r=r_{s}}\,,\nonumber\\
               &=&\pm\frac{\omega}{2\pi\alpha'}g_{xx}\sqrt{-det\,\overline{h}_{ab}}\sqrt{\overline{h}^{\tau\tau}\overline{h}^{\sigma\sigma}}\mid_{r=r_{s}}\,,\nonumber\\
               &=&-\frac{i\omega}{2\pi\alpha'}g_{xx}(r_{s})\,,
\eear
where the negative sign is chosen for the retarded function (the positivetive sign would be for the advanced function).
Therefore, the velocity dependent transverse momentum diffusion constant per unit time is given by \cite{Gursoy:2010aa}
\be\label{diff}
\kappa^\parallel(v)=-2T_{s}^\parallel\lim_{\omega\rightarrow 0}\frac{\text{Im}\,G_{R}^\parallel(\omega)}{\omega}=\frac{T_{s}^\parallel}{\pi\alpha'}g_{xx}(r_{s})=\frac{\sqrt{1+v^2}}{3\pi}\sqrt{\lambda}\mathcal{B}T\,,
\ee
where we used $g_{xx}(r_{s})=\mathcal{R}^2\mathcal{B}$, $\frac{\mathcal{R}^2}{\alpha'}=\frac{\sqrt{\lambda}}{3}$, $T_{s}^\parallel=T\sqrt{1+v^2}$.
Finally,  the jet quenching parameter $\hat{q}(v)\equiv 2\frac{\kappa^\parallel(v)}{v}$ is found to be
\be\label{qpar}
\hat{q}(v)=2\frac{\kappa^\parallel(v)}{v}=\frac{2}{3\pi}\sqrt{1+\frac{1}{v^2}}\sqrt{\lambda}\mathcal{B}T\,.
\ee

Note that when $v=0$, $\kappa^\parallel(0)$ is identified with $\kappa_\perp$, the heavy-quark momentum diffusion constant in perpendicular direction to the magnetic field introduced in Ref.\cite{Fukushima:2015wck}. Therefore, the $\mathcal{B}$ dependence of $\kappa_\perp=\frac{1}{3\pi}\sqrt{\lambda}\mathcal{B}T$ at strong coupling is similar to $\kappa_\perp\propto \alpha_{s}^2(eB)T$ found in Ref.\cite{Fukushima:2015wck} at weak coupling.

\subsubsection{$\hat q$ when the jet is perpendicular to the magnetic field}

We next consider a jet moving to $x$ direction, which is perpendicular to the magnetic field direction $z$. We first find the trailing string background as before.
Using the embedding $(\tau, \sigma)\Rightarrow (t(\tau, \sigma),x(\tau, \sigma),0,0, r=\sigma)$, and an Ansatz of the form $t(\tau,\sigma)=\tau+K(r)$ and $x=v\tau+F(r)$, the background induced metric becomes
\bear \label{bgindz2p}
\overline{h}_{\tau\tau}(v,x')&=&g_{tt}+v^2g_{xx}\,,\nonumber\\
\overline{h}_{\sigma\sigma}(v,x')&=&g_{tt}(K')^2+g_{xx}(x')^2+g_{rr}\,,\nonumber\\
\overline{h}_{\tau\sigma}(v,x')&=&g_{tt}K'+g_{xx}x'v\,.
\eear
As in the previous subsection, requiring $\overline{h}_{\tau\sigma}(v,x')=0$ to fix the residual gauge freedom, we have $\frac{\partial K}{\partial r}=-\frac{g_{xx}}{g_{tt}}x'v$ which can be used to diagonalize (\ref{bgindz2p}) as
\bear \label{bgindz3p}
\overline{h}_{\tau\tau}(v,x')&=&-g_{zz}f\Big(1-\frac{v^2}{f}\frac{g_{xx}}{g_{zz}}\Big)\,,\nonumber\\
\overline{h}_{\sigma\sigma}(v,x')&=&g_{xx}\Big(1-\frac{v^2}{f}\frac{g_{xx}}{g_{zz}}\Big)(x')^2+g_{rr}\,,\nonumber\\
\overline{h}_{\tau\sigma}(v,x')&=&0\,,
\eear
while the equation of motion becomes
\be \label{sol1p}
\frac{g_{xx}g_{zz}f \Big(1-\frac{v^2}{f}\frac{g_{xx}}{g_{zz}}\Big) x'}{\sqrt{-det\,\bar{h}_{ab}(v,x')}}={\rm constant}\equiv C_{xx}v\,.
\ee
Using $g_{rr}=\frac{1}{g_{zz}f}$ and
\be
-det\,\bar{h}_{ab}(v,x')=-\bar{h}_{\tau\tau}(v,x')\bar{h}_{\sigma\sigma}(v,x')=g_{xx}g_{zz}f\Big(1-\frac{v^2}{f}\frac{g_{xx}}{g_{zz}}\Big)^2(x')^2-(1-\frac{v^2}{f}\frac{g_{xx}}{g_{zz}})\,,
\ee
we solve (\ref{sol1p}) to obtain
\be\label{sol2p}
(x')^2=\frac{C_{xx}^2v^2}{g^2_{xx}g^2_{zz}f^2}\frac{1}{\Big(1-\frac{v^2}{f}\frac{g_{xx}}{g_{zz}}\Big)\big(1-\frac{C_{xx}^2v^2}{g_{xx}g_{zz}f}\big)}\,.
\ee
As before, the two factors in the denominator should vanish at the same location $r=\tilde{r}_{s}$, which fixes the integration constant  to be $C_{xx}=g_{xx}(\tilde{r}_s)=g_{xx}={\rm constant}$. Therefore, (\ref{sol2p}) becomes
\be\label{sol3p}
(x')^2=\frac{1}{g^2_{zz}(r)}\frac{v^2}{f^2(r)}{1\over \Big(1-\frac{v^2}{f(r)}\frac{g_{xx}}{g_{zz}(r)}\Big)^2}\,,
\ee
and using this, the metric (\ref{bgindz3p}) finally becomes
\bear \label{bgindz5p}
\overline{h}_{\tau\tau}(v,x')&=&g_{zz}\big(-f+v^2\frac{g_{xx}}{g_{zz}}\big)\,,\nonumber\\
\overline{h}_{\sigma\sigma}(v,x')&=&\frac{1}{g_{zz}\big(f-v^2\frac{g_{xx}}{g_{zz}}\big)}\,,\nonumber\\
\overline{h}_{\tau\sigma}(v,x')&=&0\,,
\eear
which can be interpreted as a 2-dimensional black hole metric with a line element $ds_{(2)}^2$ given by
\be \label{2dBH}
ds_{(2)}^{2}=\bar{h}_{\tau\tau}d\tau^2+\bar{h}_{\sigma\sigma}d\sigma^2=g_{zz}(-\tilde{\tilde{f}}(r))d\tau^2+\frac{1}{\tilde{\tilde{p}}(r)}d\sigma^2\,,
\ee
where $\tilde{\tilde{f}}(r)=f-v^2\frac{g_{xx}}{g_{zz}}$, $\tilde{\tilde{p}}(r)=g_{zz}\tilde{\tilde{f}}(r)$, and the radius of the horizon $\tilde{r}_{s}$ of the 2-dimensional black hole is found from $\tilde{\tilde{f}}(\tilde{r}_s)=0$ or $f(\tilde{r}_{s})=v^2\frac{g_{xx}}{g_{zz}(\tilde{r}_{s})}$, i.e.,
\be \label{rsperp}
\tilde{r}_{s}^2=r_{h}^2+v^2\mathcal{R}^2g_{xx}=v^2\mathcal{R}^4\mathcal{B}\Big(1+\frac{4\pi^2}{v^2}\frac{T^2}{\mathcal{B}}\Big)\,,
\ee
using $g_{xx}=\mathcal{B}\mathcal{R}^2$, $g_{zz}(\tilde{r}_{s})=\frac{\tilde{r}_{s}^2}{\mathcal{R}^2}$, and $T=\frac{r_{h}}{2\pi \mathcal{R}^2}$ from (\ref{TBTZ}).
The Hawking temperature of this 2-dimensional black hole is given by
\be \label{Tsperp}
T_{s}^{\perp}=\frac{1}{4\pi}\sqrt{g_{zz}(\tilde{r}_{s})\tilde{\tilde{f}}'(\tilde{r}_{s})\tilde{\tilde{p}}'(\tilde{r}_{s})}=\frac{g_{zz}(\tilde{r}_{s})\tilde{\tilde{f}}'(\tilde{r}_{s})}{4\pi}=\frac{v\sqrt{\mathcal{B}}}{2\pi}\Big(1+\frac{4\pi^2}{v^2}\frac{T^2}{\mathcal{B}}\Big)^{1/2}\,.
\ee

Note that the drag force to the heavy-quark jet $F_{drag}^{\perp}$ is simply given by
\be\label{dragx}
F_{drag}^{\perp}=\frac{\delta\mathcal{L}}{\delta x'}=-\dfrac{C_{xx}}{2\pi\alpha'}v=-\dfrac{1}{6\pi}\sqrt{\lambda}\mathcal{B}v\,,
\ee
where we used $C_{xx}=g_{xx}$ in the last equality.
It is interesting to note that this drag force exists even at zero temperature. As we explained in the case of weak coupling, this is possible in the case of weak coupling due to the fact that it is kinematically possible to create a quark-antiquark pair from the LLL vacuum by scatterings with the jet. It is interesting that we observe the same feature even at strong coupling.

To find the transverse momentum diffusion along $z$ direction (note that $z,y$ are the two perpendicular directions to the jet motion),
we consider fluctuations of string position along the $z$ direction which is dual to the $z$ component of color electric field: $\delta z$.
The transverse fluctuation $\delta h_{ab}(\delta \dot{z},\delta z')$ around the background induced metric $\overline{h}_{ab}(v,x')$ (\ref{bgindz5p}) is given by
\bear \label{flucz1p}
\delta h_{\tau\tau}(\delta \dot{z},\delta z')&=&g_{zz}(\delta \dot{z})^2\,,\nonumber\\
\delta h_{\sigma\sigma}(\delta \dot{z},\delta z')&=&g_{zz}(\delta z')^2\,,\nonumber\\
\delta h_{\tau\sigma}(\delta \dot{z},\delta z')&=&g_{zz}(\delta \dot{z}\delta z')^2\,,
\eear
 and the Nambu-Goto action is expanded to linear order in $\delta h_{ab}(\delta \dot{z},\delta z')$ as
\bear \label{NGlinear1p}
S^\perp_{NG}&=&\int d\tau d\sigma \mathcal{L}^\perp\big(\overline{h}_{ab}(v,x'),\delta h_{ab}(\delta \dot{z},\delta z')\big)\,,\nonumber\\
                &=&-\frac{1}{4\pi\alpha'}\int d\tau d\sigma g_{zz}\sqrt{-det\,\overline{h}_{ab}(v,x')}\overline{h}^{ab}(v,x')\delta h_{ab}(\delta \dot{z},\delta z')\,,\nonumber\\
                &=&-\frac{1}{2}\int d\tau d\sigma \overline{G}_\perp^{ab}(v,x')\partial_{a}\delta z(\tau,\sigma)\partial_{b}\delta z(\tau,\sigma)\,,
\eear
where $\overline{G}_\perp^{ab}(v,x')\equiv\frac{1}{2\pi\alpha'}g_{zz}\sqrt{-det\,\overline{h}_{ab}(v,x')}\overline{h}^{ab}(v,x')$.
Using the conjugate momenta $\Pi^\perp=\frac{\partial\mathcal{L}^\perp}{\partial_{\sigma}\delta z}$, defining the retarded Green's function as $G_{R}^\perp\equiv-\frac{\Pi^\perp}{\delta z}$, and using the equation of motion for $\delta z$ in momentum space derived from the action (\ref{NGlinear1p})
\be\label{EOM2}
\partial_{\sigma}\overline{G}_\perp^{\sigma\sigma}\partial_{\sigma}\delta z-\omega^2\overline{G}_\perp^{\tau\tau}\delta z=0\,,
\ee
one can derive the holographic RG flow equation for the retarded Green's function $G_{R}^\perp$ to be
\be\label{RG}
\partial_{\sigma}G_{R}^\perp=-\frac{(G_{R}^\perp)^2}{\overline{G}_\perp^{\sigma\sigma}}+\omega^2\overline{G}_\perp^{\tau\tau}.
\ee
By the same reasoning as before, we have in small $\omega$ limit
\bear\label{GR}
G_{R}^\perp(\omega)&=&\pm\omega\sqrt{\overline{G}_\perp^{\tau\tau}\overline{G}_\perp^{\sigma\sigma}}\mid_{r=r_{s}}\,,\nonumber\\
               &=&\pm\frac{\omega}{2\pi\alpha'}g_{zz}\sqrt{-det\,\overline{h}_{ab}}\sqrt{\overline{h}^{\tau\tau}\overline{h}^{\sigma\sigma}}\mid_{r=\tilde{r}_{s}}\,,\nonumber\\
               &=&-\frac{i\omega}{2\pi\alpha'}g_{zz}(\tilde{r}_{s})\,.
\eear

Therefore, the velocity dependent momentum diffusion constant along $z$ per unit time when the jet is moving perpendicular to the magnetic field is given by
\be\label{diffp}
\kappa_{z}^\perp(v)=-2T_{s}^\perp\lim_{\omega\rightarrow 0}\frac{\text{Im}\,G_{R}^\perp(\omega)}{\omega}=\frac{T_{s}^\perp}{\pi\alpha'}g_{zz}(\tilde{r}_{s})=\frac{v^{3}}{6\pi^2}\sqrt{\lambda}\mathcal{B}^{3/2}+v\sqrt{\lambda}\sqrt{\mathcal{B}}T^2\,,
\ee
for $\mathcal{B}\gg T^2$, where we have used $g_{zz}(\tilde{r}_{s})=\frac{\tilde{r}_{s}^2}{\mathcal{R}^2}=v^2\mathcal{R}^2\mathcal{B}\big(1+\frac{4\pi^2}{v^2}\frac{T^2}{\mathcal{B}}\big)$ from (\ref{rsperp}), $\frac{\mathcal{R}^2}{\alpha'}=\frac{\sqrt{\lambda}}{3}$, $T_{s}^\perp=\frac{v\sqrt{\mathcal{B}}}{2\pi}\Big(1+\frac{4\pi^2}{v^2}\frac{T^2}{\mathcal{B}}\Big)^{1/2}$ from (\ref{Tsperp}).
Therefore, the jet quenching parameter $\hat{q}_{z}\equiv \frac{\kappa_{z}^\perp(v)}{v}$ is given by
\be\label{qzperp}
\hat{q}_{z}=\frac{\kappa_{z}^\perp(v)}{v}=\frac{v^{2}}{6\pi^2}\sqrt{\lambda}\mathcal{B}^{3/2}+\sqrt{\lambda}\sqrt{\mathcal{B}}T^2\,,
\ee
which has a very similar structure to that at weak coupling (\ref{weaksum}). Especially, the first term is the vacuum part that exists even at zero temperature, similarly to the case at weak coupling.

Note that when $v=0$, $g_{zz}(\tilde{r}_{s})=g_{zz}(r_{h})=4\pi^2\mathcal{R}^2T^2$, and $\kappa_{z}^\perp(0)=\frac{T}{\pi\alpha'}g_{zz}(r_{h})=\frac{4\pi}{3}\sqrt{\lambda}T^3$ is identified with $\kappa_\parallel$, the heavy-quark momentum diffusion constant along the magnetic field introduced in Ref.\cite{Fukushima:2015wck}. Therefore $\kappa_\parallel$ at strong coupling is independent of $\mathcal{B}$, which is precisely same to $\kappa_\parallel\propto\alpha_{s}^2T^3$ in Ref.\cite{Fukushima:2015wck} found at weak coupling pQCD. This seems in line with the idea of superfluid nature of LLL states in Ref.\cite{Sadofyev:2015tmb}.

Following the same steps, one can compute the diffusion constant along the other remaining transverse direction $y$.
We find the momentum diffusion per unit time as
\be\label{diff}
\kappa_{y}^\perp(v)=\frac{T_{s}^\perp}{\pi\alpha'}g_{yy}(\tilde{r}_{s})=\frac{T_{s}^\perp}{\pi\alpha'}g_{xx}=\frac{v}{6\pi^2}\sqrt{\lambda}\mathcal{B}^{3/2}+\dfrac{1}{3v}\sqrt{\lambda}\sqrt{\mathcal{B}}T^2\,,
\ee
for $\mathcal{B}\gg T^2$, and we finally have
\be\label{qyperp}
\hat{q}_{y}=\frac{\kappa_{y}^\perp(v)}{v}=\frac{1}{6\pi^2}\sqrt{\lambda}\mathcal{B}^{3/2}+\dfrac{1}{3v^2}\sqrt{\lambda}\sqrt{\mathcal{B}}T^2\,.
\ee
It is interesting to compare these results in AdS/CFT, (\ref{qzperp}) and (\ref{qyperp}), with the results at weak coupling (\ref{weaksum}) computed in pQCD.

\subsection{$\hat q$ from light-like Wilson loop\label{sec32}}

The jet quenching parameter at strong coupling was first computed in Ref.\cite{Liu:2006ug, Liu:2006he} at zero magnetic field using light-like Wilson loops and the AdS/CFT correspondence. See also Ref.\cite{Panero:2013pla} for the lattice QCD computation of the jet quenching parameter. Here, we extend the works of Ref.\cite{Liu:2006ug, Liu:2006he} to the case with strong magnetic field by using the general formula for jet quenching parameter derived in Ref.\cite{Giataganas:2012zy}. Since the computational steps are already in literature, we simply summarize the general formula and our results in the case of strong magnetic field. We emphasize that the results we obtain from this method
are different from those we obtain in the previous subsection using the heavy-quark trailing string: this discrepancy exists even in the original computations for ${\cal N}=4$ SYM without magnetic field. This might be due to possible breakdown of heavy-quark method in ultra-relativistic limit \cite{CasalderreySolana:2007qw}, although it has not been fully understood to the best of our knowledge.

\subsubsection{$\hat q$ when the jet is parallel to the magnetic field}

We first make a coordinate transformation $r=\dfrac{\mathcal{R}^2}{u}$ to rewrite our metric (\ref{BTZ}) as
\be \label{BTZu}
ds^{2}=G_{\mu\nu}dx^{\mu}dx^{\nu}=\frac{\mathcal{R}^2}{u^2}\left(-f(u)dt^2+dz^2\right)+\mathcal{R}^2\mathcal{B}(dx^2+dy^2)+\frac{\mathcal{R}^2}{u^2f(u)}du^2\,,
\ee
where $f(u)=1-\frac{u^2}{u_{h}^2}$, the horizon corresponds to $u=u_{h}$, the boundary to $u=0$, and the Hawking temperature $T$ of the BTZ black hole (\ref{BTZu}) is
\be \label{TBTZu}
T=\frac{1}{2\pi u_{h}}.
\ee

The jet quenching parameter $\hat{q}$ for a jet moving along the $z$ direction (with the speed of light $v=1$) can be directly obtained from the metric by \cite{Giataganas:2012zy},
\be\label{qzx1}
\hat{q}=\frac{1}{\pi\alpha'}\Bigg(\int_{0}^{u_{h}}du\frac{1}{G_{xx}}\sqrt{\dfrac{G_{uu}}{G_{tt}+G_{zz}}}\Bigg)^{-1}=\frac{2}{3}\sqrt{\lambda}\mathcal{B}T\Bigg(\int_{0}^{u_{h}}du \sqrt{1\over u^2-u^4/u_h^2}\Bigg)^{-1}\,.
\ee
The integral in the above has a logarithmic UV divergence near $u=0$, which is easy to understand. Recall that our BTZ metric (\ref{BTZu})
is valid only up to the ``UV cutoff'' $u_c\approx 1/\sqrt{{\cal B}}$ in the full 5 dimensional dual geometry where the energy scale $1/u$ is smaller than the scale of the magnetic field. For $u\ll u_c$, especially near the UV boundary $u=0$, the full $AdS_5$ geometry takes over, which makes the above integral finite in the region $u\ll u_c$.  Therefore, a large logarithm develops in the above integral between the scale $1/u_h\sim T$ and $1/u_c\sim \sqrt{\cal B}$, and we get the leading-log result of $\hat q$ as
\be
\hat q=\frac{4}{3}{\sqrt{\lambda}\mathcal{B}T\over \log({\cal B}/T^2)}\,.
\ee
To find the constant under the log, we need to know the exact geometry interpolating BTZ and $AdS_5$, but we will not
go into such detail in this work, satisfied with the above leading-log result in our assumed hierarchy ${\cal B}\gg T^2$.

\subsubsection{$\hat q$ when the jet is perpendicular to the magnetic field }

We compute next the jet quenching parameter when the jet is moving perpendicular to the magnetic field. As we can have two different transverse directions, one along the magnetic field, the other perpendicular to the magnetic field, we should consider the two cases separately as before. Let the magnetic field point to $z$ direction, and the jet move to $x$ direction.

The jet quenching parameter $\hat{q}_{z}$ for the momentum broadening along the $z$ direction is
\be\label{qxz1}
\hat{q}_{z}=\frac{1}{2\pi\alpha'}\Bigg(\int_{u_c}^{u_{h}}du\frac{1}{G_{zz}}\sqrt{\dfrac{G_{uu}}{G_{tt}+G_{xx}}}\Bigg)^{-1}=\frac{\sqrt{\lambda}}{6\pi}\Bigg(\int_{u_c}^{u_{h}}du\frac{u^2}{\sqrt{(\mathcal{B}u^2-1+u^2/u_h^2)(1-u^2/u_h^2)}}\Bigg)^{-1}\,,
\ee
where an extra factor $1/2$ is from our definition of $\hat q_z$ (such that in an isotropic case, $\hat q=\hat q_z+\hat q_y=2\hat q_z$),
and $u_c\approx 1/\sqrt{{\cal B}}$ is the UV cutoff of our BTZ metric. From the above, it is easy to see that the region $u\lesssim u_c$ gives a contribution to the integral which is of order $u_c^3\sim (1/{\cal B})^{3/2}$, that is subleading compared to the contribution from $u_c\ll u<u_h$, where the integral becomes simplified to
\be
\int_{u_c}^{u_{h}}du\frac{u^2}{\sqrt{(\mathcal{B}u^2-1+u^2/u_h^2)(1-u^2/u_h^2)}}\approx {1\over \sqrt{\cal B}} \int_{0}^{u_{h}}du\frac{u}{\sqrt{1-u^2/u_h^2}}={u_h^2\over \sqrt{{\cal B}}}={1\over 4\pi^2 \sqrt{{\cal B}}T^2}\,,
\ee
so that we have a leading order expression for $\hat q_z$ as
\be\label{qxz2}
\hat{q}_{z}=\frac{2\pi}{3}\sqrt{\lambda}\sqrt{\mathcal{B}}T^2\,.
\ee

Similarly,
the momentum broadening along $y$ direction, $\hat{q}_{y}$, is
\be\label{qxy1}
\hat{q}_{y}=\frac{1}{2\pi\alpha'}\Bigg(\int_{u_c}^{u_{h}}du\frac{1}{G_{yy}}\sqrt{\dfrac{G_{uu}}{G_{tt}+G_{xx}}}\Bigg)^{-1}=\frac{\sqrt{\lambda}\mathcal{B}}{6\pi}\Bigg(\int_{u_c}^{u_{h}}du\frac{1}{\sqrt{(\mathcal{B}u^2-1+u^2/u_h^2)(1-u^2/u_h^2)}}\Bigg)^{-1}\,.
\ee
The integral produces a leading large logarithm between $u_c\ll u\ll u_h$ where the integral becomes
\be
\int_{u_c}^{u_{h}}du\frac{1}{\sqrt{(\mathcal{B}u^2-1+u^2/u_h^2)(1-u^2/u_h^2)}}\approx {1\over\sqrt{\cal B}}\int_{u_c}^{u_{h}}du\frac{1}{u}={1\over 2\sqrt{\cal B}}\log({\cal B}/T^2)\,.
\ee
The constant under the log requires a full knowledge of the interpolating metric between BTZ and $AdS_5$, and it is easy to see that the UV region $u\lesssim u_c$ also produces a constant under the log. Therefore, we have a leading-log result for $\hat q_y$ in ${\cal B}\gg T^2$ limit as
\be
\hat q_y={\sqrt{\lambda}{\cal B}^{3/2}\over (3\pi)\log({\cal B}/T^2)}\,.
\ee
Comparing with (\ref{qxz2}), we see that $\hat q_y\gg \hat q_z$ in the assumed hierarchy ${\cal B}\gg T^2$.

 \section{Summary and Discussion \label{summary}}

 Our results are summarized as follows. In weak coupling perturbative QCD, for a jet moving parallel to the strong magnetic field, we have the jet quenching parameter at complete leading order in $\alpha_s$ (the leading log and the constant under the log) as
 \be
 \hat q={1\over\pi}(1+1/v)C_2^JT_RN_F\,\alpha_s^2  (eB) T\Big(\log\left(1/\alpha_s\right)-1-\gamma_E-\log\left(T_RN_F/\pi\right)\Big)\,.
 \ee
 For a jet moving perpendicular to the magnetic field, there are two different transverse directions due to the presence of the magnetic field. The momentum diffusion along the magnetic field direction, $\hat q_z$, is given by
 \be\label{qzs}
 \hat q_z={16 v^2\over 3(2\pi)^{3/2}} C_2^J T_RN_F \alpha_s^2 (eB)^{3/2}+{1\over\pi}{v} C_2^JT_R N_F \alpha_s^2(eB)T\left(\log\left(T^2\over \alpha_s T_RN_F\left(eB\over 2\pi\right) v^2\right)-2\right)\,,
 \ee
 while the momentum diffusion along the perpendicular direction, $\hat q_y$, is given by
 \be\label{qys}
 \hat q_y={8 \over 3(2\pi)^{3/2}} C_2^J T_RN_F \alpha_s^2 (eB)^{3/2}+{1\over\pi v} C_2^JT_R N_F \alpha_s^2(eB)T\left(\log\left(T^2\over \alpha_s T_RN_F\left(eB\over 2\pi\right) v^2\right)+0\right)\,.
 \ee
 In both (\ref{qzs}) and (\ref{qys}), the first term represents the vacuum contribution that exists even at zero temperature, while the second term is the leading thermal contribution to complete leading order (the leading log and the constant under the log). These two terms are the first two leading contributions in the assumed hierarchy of scales, $\alpha_s eB \ll T^2\ll eB$.

 In strong coupling AdS/CFT correspondence, we compute our jet quenching parameters in the two different methods: 1) heavy-quark strings, and 2) light-like Wilson loops.
 In the method 1), when the jet is moving parallel to the magnetic field, we have ($\lambda\equiv g_s^2 N_c$ and ${\cal B}=eB$)
 \be
 \hat q=\frac{2}{3\pi}\sqrt{1+\frac{1}{v^2}}\sqrt{\lambda}\mathcal{B}T\,,
 \ee
 while, in the case the jet is moving perpendicular to the magnetic field, the two different momentum diffusion constants depending on the orientation with respect to magnetic field are
 \be
 \hat q_z=\frac{v^{2}}{6\pi^2}\sqrt{\lambda}\mathcal{B}^{3/2}+\sqrt{\lambda}\sqrt{\mathcal{B}}T^2\,,
 \ee
 and
 \be
 \hat q_y=\frac{1}{6\pi^2}\sqrt{\lambda}\mathcal{B}^{3/2}+\dfrac{1}{3v^2}\sqrt{\lambda}\sqrt{\mathcal{B}}T^2\,.
 \ee

 In the method 2) of the AdS/CFT correspondence, for the jet moving parallel to the magnetic field, we have
 \be
 \hat q=\frac{4}{3}{\sqrt{\lambda}\mathcal{B}T\over \log({\cal B}/T^2)}\,,
 \ee
 and for the jet moving perpendicular to the magnetic field, we have
 \be
 \hat q_z=\frac{2\pi}{3}\sqrt{\lambda}\sqrt{\mathcal{B}}T^2\,,
 \ee
 and
 \be\hat q_y={\sqrt{\lambda}{\cal B}^{3/2}\over (3\pi)\log({\cal B}/T^2)}\,.
 \ee

 Perhaps, the most useful observations from these results in the assumed hierarchy $T^2\ll eB$ are
 1) the jet quenching is generally larger in the case the jet is moving perpendicular to the magnetic field, compared to the case the jet is moving parallel to the magnetic field, 2) in the case the jet is moving perpendicular to the magnetic field, the transverse momentum diffusion is asymmetric, $\hat q_z\neq \hat q_y$. The 1) implies that the strong magnetic field tends to suppress more jets in the reaction plane than the jets out-of reaction plane, so it would reduce the elliptic flow of the jets.
The 2) implies that the BDMPS-Z/LPM evolution equation of the gluon emission vertex $F(\bm b)$ in the two dimensional impact parameter space $\bm b$ in large scattering number limit (that is, small $\bm b$ limit, or harmonic potential limit) becomes an asymmetric harmonic oscillator problem with complex frequencies,
\be
i{\partial F(\bm b)\over\partial t}=-{1\over 2\omega}\nabla^2_{\bm b} F(\bm b) +{i\over 4}\left(\hat q_z {\bm b}_z^2+\hat q_y {\bm b}_y^2\right) F(\bm b)\,,
\ee
 where ${\bm b}=({\bm b}_z,{\bm b}_y)$ and $\omega$ is the gluon energy. This problem is still solvable analytically both in finite and infinite mediums, which can be plugged into the emission formula to find the azimuthally asymmetric gluon Bremsstrahlung spectrum.
We hope to report a detailed numerical analysis of it and its implications in heavy-ion phenomenology of jet spectrum in a near future.

 \vskip 1cm \centerline{\large \bf Acknowledgment} \vskip 0.5cm

 We thank Tigran Kalaydzhyan, Rob Pisarski, Andrey Sadofyev, Misha Stephanov for discussions.
 We appreciate the hospitality and support provided by RIKEN-BNL Research Center. This material is based upon work partially supported by the U.S. Department of Energy, Office of Science, Office of Nuclear Physics, within the framework of the Beam Energy Scan Theory (BEST) Topical Collaboration.

\section*{Appendix: Thermal gluon contribution in leading log }

In this appendix, we compute the contribution to the jet quenching parameter coming from the scatterings with thermal gluons (rather than scatterings with thermal LLL states in the main text). The purpose is to confirm our statement in the introduction that this contribution is of order
\be
\hat q_{\rm gluon}\sim \alpha_s^2T^3  \log\left( T^2\over\alpha_s eB\right)\,,
\ee
which is indeed subleading compared to the LLL scattering contributions in the main text which is of order
\be
\hat q_{\rm quark}\sim \alpha_s^2 (eB)T\log(1/\alpha_s)\quad {\rm or}\quad \alpha_s^2 (eB)T \log\left( T^2\over\alpha_s eB\right)\,,
\ee
in our assumed hierarchy of scales $\alpha_s eB\ll T^2\ll eB$. For simplicity, we will present our computation only for the case where the jet is moving parallel to the magnetic field.

The starting point is the expression for the scattering rate with a momentum transfer $\bm q$ in large jet momentum limit $P\to \infty$:
\bear
{d\Gamma\over d^3\bm q}&=&{1\over (2 E_{\bm p})^2 (2\pi)^3}\int {d^3\bm k\over (2\pi)^3 2|\bm k|}{1\over 2|\bm k-\bm q|}\big|{\cal M}\big|^2 n_B(|\bm k|)\left(n_B(|\bm k-\bm q|)+1\right)\nonumber\\&\times&(2\pi)\delta\left(|\bm k|-|\bm k-\bm q|-\bm v\cdot\bm q\right)\,,\label{sct1}
\eear
where $\bm k$ is the momentum of the incoming gluon of polarization $\epsilon$, and $\bm k-\bm q\equiv\bm k'$ is the momentum of the out-going gluon of polarization $\epsilon'$, and the last $\delta$-function is from the energy conservation,
\be
q^0\equiv|\bm k|-|\bm k-\bm q|=E_{\bm p+\bm q}-E_{\bm p}\approx \bm v\cdot\bm q\,,\label{ec}
\ee
where we used
\be
E_{\bm p+\bm q}-E_{\bm p}\approx {\partial E_{\bm p}\over\partial\bm p}\cdot \bm q={\bm p\over E_{\bm p}}\cdot\bm q=\bm v\cdot \bm q\,.
\ee

\begin{figure}[t]
 \centering
 \includegraphics[height=7cm]{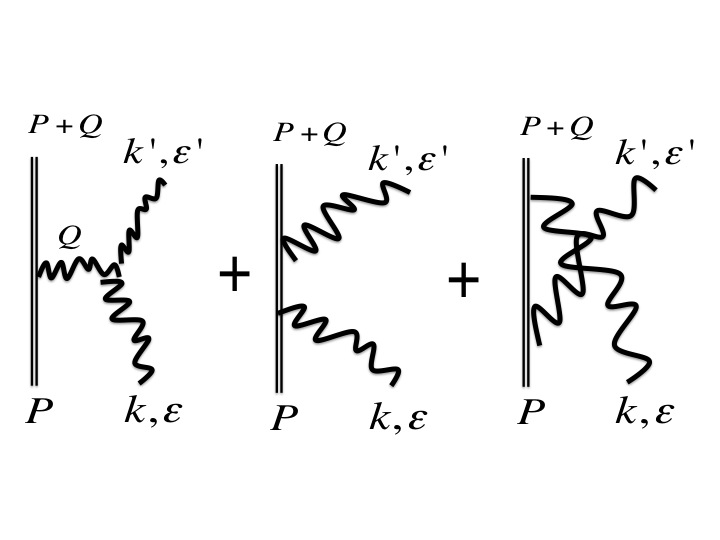}\caption{Three Feynman diagrams for $\cal M$. The first diagram dominates in large $P$ limit in the gauge $\epsilon\cdot P=\epsilon'\cdot P=0$. \label{apfig1}}
 \end{figure}

There are three diagrams contributing to the matrix element $\cal M$ as shown in Figure \ref{apfig1}.
In general we need all three to have a gauge invariant amplitude. We are interested in the leading large $P$ limit of the amplitude,
and in this case, it has been known (see Ref.\cite{Moore:2004tg}) that only the first $t$-channel diagram gives the leading $P$ result if we work in the special gauge $\epsilon\cdot P=\epsilon'\cdot P=0$. From this we have
\bear
{\cal M}&=&gf^{abc}\,\bar U(P+Q) ig t^a\gamma^\beta U(P) G_{\beta\mu}(Q) \nonumber\\&\times&\left((-Q_\nu+k_\nu')\eta_{\mu\alpha}
+(k_\alpha+Q_\alpha)\eta_{\mu\nu}-(k_\mu+k_\mu')\eta_{\alpha\nu}\right) \epsilon^\nu(\tilde\epsilon^\alpha)^*\,,\label{calM}
\eear
where $b$ and $c$ denote the color charges of the incoming and out-going thermal gluons, while $G_{\beta\mu}(Q)$
 is the t-channel gluon propagator for the exchanged gluon line. The Debye screening provided by the LLL states is
 included in $G_{\beta\mu}(Q)$ by re-summing the gluon self-energy in this propagator, and it is given previously by (\ref{gra}) which we repeat here
\be
G_{\beta\mu}(Q)=  -i{\eta_{\beta\mu}\over Q^2}+i(1-\xi){Q_\beta Q_\mu\over (Q^2)^2}
-\left(Q_\parallel^2 \eta_{\parallel\beta\mu}-Q_{\parallel\beta}Q_{\parallel\mu}\right){\chi\over Q^2(Q^2+i\chi Q_\parallel^2)}\,,
 \label{graapp}
\ee
with
\be
\chi\equiv -4i\alpha_s  T_R N_F \left(eB\over 2\pi\right) e^{-{{\bm q}_\perp^2\over 2eB}} {1\over Q_{\parallel}^2}\,.
\ee
The color trace gives a factor $C_R N_c$, and in large $P$ limit, we have $U(P+Q) \gamma^\beta U(P)\approx 2P^\beta$.
Due to the on-shell Ward identity, the gauge dependent second term in (\ref{graapp}) does not contribute, and by the same reason we can replace $Q_{\parallel\beta}$ with $-Q_{\perp\beta}$ in the last term. Then, since $P^\beta\cdot Q_{\perp\beta}=0$ this last term is subleading in large $P$ limit. Therefore we have
\be
P^\beta G_{\beta\mu}(Q)\approx -i{P_\mu\over Q^2}-{P_\mu \,\chi Q_{\parallel}^2\over Q^2(Q^2+i\chi Q_\parallel^2)}=-i {P_\mu\over
Q^2+m_{D,B}^2}\,,
\ee
where the effective Debye mass is given by
\be
m_{D,B}^2\equiv i\chi Q_{\parallel}^2=4\alpha_s T_R N_F \left(eB\over 2\pi\right) e^{-{{\bm q}_\perp^2\over 2eB}}\,,\label{dbye}
\ee
and using the gauge $\epsilon\cdot P=\tilde\epsilon'\cdot P=0$, we have from (\ref{calM}) (neglecting colors)
\be
{\cal M}\approx {-2 g^2\over (Q^2+m_{D,B}^2)}(P\cdot(k+k'))\left(\epsilon\cdot\tilde\epsilon\right)\,.
\ee

The rest of the computation parallels the steps in Ref.\cite{Moore:2004tg}.
Using the energy conservation (\ref{ec}) written covariantly as $P\cdot Q=0$ we have $P\cdot(k+k')=2(P\cdot k)$,
and the polarization sum in our gauge is worked out as in Ref.\cite{Moore:2004tg} to give
\be
\sum_{\epsilon,\tilde\epsilon}|\epsilon\cdot\tilde\epsilon'|^2=2-{P^2 Q^2\over (P\cdot Q)^2}+{(Q^2)^2\over 4}{(P^2)^2\over (P\cdot k)^4}\,,
\ee
so that
\be
\big|{\cal M}\big|^2=C_R N_c {16 g^4\over (Q^2+m_{D,B}^2)^2}\left(2(P\cdot k)^2-Q^2 P^2+{(Q^2)^2\over 4}{(P^2)^2\over (P\cdot k)^2}\right)\,.\label{mapp}
\ee

We compute the leading log contribution to $\hat q$ from the scattering rate (\ref{sct1}) with (\ref{mapp}).
Since we expect that the leading log comes from the range $m_{D,B}^2\ll Q^2\ll T^2$ while $k\sim T$ (due to $n_B(|\bm k|)$ in (\ref{sct1})), we can simplify things in the approximation $Q\ll k$:
\be
q^0=|\bm k|-|\bm k-\bm q|\approx \hat{\bm k}\cdot\bm q\,,
\ee
and
\be
\big|{\cal M}\big|^2\approx C_R N_c {32 g^4\over (Q^2+m_{D,B}^2)^2}(P\cdot k)^2\,,
\ee
with $P\cdot k=-E_{\bm p}(|\bm k|-\bm v\cdot\bm k)$.
To perform angular integrations, we choose the direction of $\bm q$ to be along $\hat z$ and let $\bm p$ lie on $(x,z)$ plane making an angle $\theta_{\bm q}$ from $\bm q$, and introduce spherical coordinate $(\theta,\phi)$ for $\bm k$. Then, the energy $\delta$-function becomes
\be
\delta(\hat{\bm k}\cdot\bm q-\bm v\cdot\bm q)={1\over |\bm q|}\delta(\cos\theta-v\cos\theta_{\bm q})\,,
\ee
and
\be
P\cdot k=-E_{\bm p}(|\bm k|-\bm v\cdot\bm k)=-E_{\bm p}|\bm k|\left(1-v(\sin\theta_{\bm q}\sin\theta\cos\phi+\cos\theta_{\bm q}\cos\theta)\right)\,,
\ee
so we have
\bear
{d\Gamma\over d^3\bm q}&\approx& {2 C_RN_c g^4\over (2\pi)^5 |\bm q|}\int_0^\infty d|\bm k| |\bm k|^2 n_B(|\bm k|)(n_B(|\bm k|)+1)\nonumber\\
&\times&\int^1_{-1}d\cos\theta \int_0^{2\pi}d\phi{ \left(1-v(\sin\theta_{\bm q}\sin\theta\cos\phi+\cos\theta_{\bm q}\cos\theta)\right)^2\over \left(|\bm q|^2(1-\cos^2\theta)+m_{D,B}^2\right)^2}\delta(\cos\theta-v\cos\theta_{\bm q})\nonumber\\&=&
{C_R N_c g^4 T^3\over 6(2\pi)^2}{1\over |\bm q|}{1\over \left(|\bm q|^2(1-v^2\cos^2\theta_{\bm q})+m_{D,B}^2\right)^2}\nonumber\\&\times&
\left(1-2v^2\cos^2\theta_{\bm q}+v^4\cos^4\theta_{\bm q}+{1\over 2}v^2\sin^2\theta_{\bm q}(1-v^2\cos^2\theta_{\bm q})\right)\,,
\eear
where we used
\be
\int_0^\infty d|\bm k| |\bm k|^2 n_B(|\bm k|)(n_B(|\bm k|)+1)={\pi^2 T^3\over 3}\,,
\ee
and $m_{D,B}^2\sim \alpha_s eB$ here is a constant dropping exponential factor in (\ref{dbye}) since $Q^2\ll T^2\ll eB$.

Since ${\bm q}_\perp^2=|\bm q|^2\sin^2\theta_{\bm q}$, we finally obtain the jet quenching parameter up to leading log in $T^2/m_{D,B}^2$ as
\bear
\hat q_{\rm gluon}&=&{1\over v}\int d^3\bm q {d\Gamma\over d^3\bm q} {\bm q}_{\perp}^2\nonumber\\
&=&{C_RN_c g^4 T^3\over 6(2\pi)v}\int d\cos\theta_{\bm q}\int_0^{T}d |\bm q| {|\bm q|^3\sin^2\theta_{\bm q}\over
 \left(|\bm q|^2(1-v^2\cos^2\theta_{\bm q})+m_{D,B}^2\right)^2}\nonumber \\ &\times&
 \left(1-2v^2\cos^2\theta_{\bm q}+v^4\cos^4\theta_{\bm q}+{1\over 2}v^2\sin^2\theta_{\bm q}(1-v^2\cos^2\theta_{\bm q})\right)\nonumber\\
 &\approx&{C_RN_c g^4 T^3\over 12(2\pi)v}\log\left(T^2\over \alpha_s eB\right)\int d\cos\theta_{\bm q} {\sin^2\theta_{\bm q}\over (1-v^2\cos^2\theta_{\bm q})^2}\nonumber \\ &\times&
 \left(1-2v^2\cos^2\theta_{\bm q}+v^4\cos^4\theta_{\bm q}+{1\over 2}v^2\sin^2\theta_{\bm q}(1-v^2\cos^2\theta_{\bm q})\right)\nonumber\\
 &=&{C_RN_c g^4 T^3\over 12(2\pi)}\log\left(T^2\over \alpha_s eB\right)\left({3\over v}-{1\over v^3}+{(1-v^2)^2\over 2 v^4}\log\left(1+v\over 1-v\right)\right)\,.
\eear
Up to the change $m_D^2\sim\alpha_s T^2\to m^2_{D,B}\sim\alpha_s eB$ inside the logarithm, the velocity dependence in this leading log expression is identical to the case without magnetic field obtained in Ref.\cite{Moore:2004tg}.

\vfil

\end{document}